\documentclass[11pt,onecolumn,twoside]{IEEEtran}
\usepackage{amsmath, amssymb, graphicx, cite}
\usepackage{algorithmic}

\newtheorem{theorem}{Theorem}
\newtheorem{lemma}{Lemma}
\newtheorem{proposition}{Proposition}
\newtheorem{corollary}{Corollary}

\newtheorem{definition}{Definition}

\newcommand{\be}{\begin{eqnarray}}
\newcommand{\ee}{\end{eqnarray}}

\newcommand{\EX}{\mathbf{E}}
\newcommand{\PR}{\mathbf{P}}
\newcommand{\Z}{\mathbb{Z}}

\newcommand{\mbf}{\mathbf}
\newcommand{\geo}{\mathsf{geo}}

\begin{document}

\title{Capacity of Systems with Queue-Length Dependent Service Quality}

\author{Avhishek~Chatterjee,
	Daewon~Seo, and
 	Lav~R.~Varshney
\thanks{The authors are with the Coordinated Science Laboratory, University of Illinois at Urbana-Champaign, Urbana, IL 61801, USA 
(e-mail: \{avhishek, dseo9, varshney\}@illinois.edu). This work was supported in part by the National Science Foundation under grant CCF-1623821. }
}

\maketitle

\begin{abstract}
We study the information-theoretic limit of reliable information processing by a server with queue-length dependent quality of service. We define the capacity for such a system as the number of bits reliably processed per unit time, and characterize it in terms of queuing system parameters. We also characterize the distributions of the arrival and service processes that maximize and minimize the capacity of such systems in a discrete-time setting. For arrival processes with at most one arrival per time slot, we observed a minimum around the memoryless distribution. We also studied the case of multiple arrivals per time slot, and observed that burstiness in arrival has adverse effects on the system. The problem is theoretically motivated by an effort to incorporate the notion of reliability in queueing systems, and is applicable in the contexts of crowdsourcing, multimedia communication, and stream computing.
\end{abstract}


\section{Introduction}
\label{sec:intro}
Consider the following abstraction of a centrally-controlled system of jobs and servers: a job requester places a request to a central controller for a server to process a large group of jobs. The central controller, considering factors such as availability of servers and commitment to other customers, chooses an appropriate server. Then it gradually dispatches jobs to that server. The server processes the jobs as they arrive, following a first-in first-out queuing discipline. This models many systems with jobs and servers, such as in crowdsourcing, multimedia communication, and stream computing.

In addition to the queuing discipline, we consider the server to be imperfect. The quality of service received by a job depends on the number of jobs waiting in the queue at that time: longer queues lead to more noise in information processing. The overall performance of the server on the group of jobs, therefore, depends on the state of the queue as it evolves over the entire service duration. Hence, in turn, it depends on the service requirements of the jobs, their arrivals  to the server or the dispatch time from the controller, and the relation between service quality and queue state. If the job requester knows the dispatch and service distributions, it can increase overall reliability by designing the group of jobs, either with added redundancy or combining them appropriately. In this work, we formally characterize the maximum rate at which jobs can be reliably processed in such a requester-dispatcher-server system and study various queuing systems and their parameters, e.g., arrival and service distributions, to achieve optimal rates.

\begin{figure}
  \centering
  \includegraphics[width=3.5in]{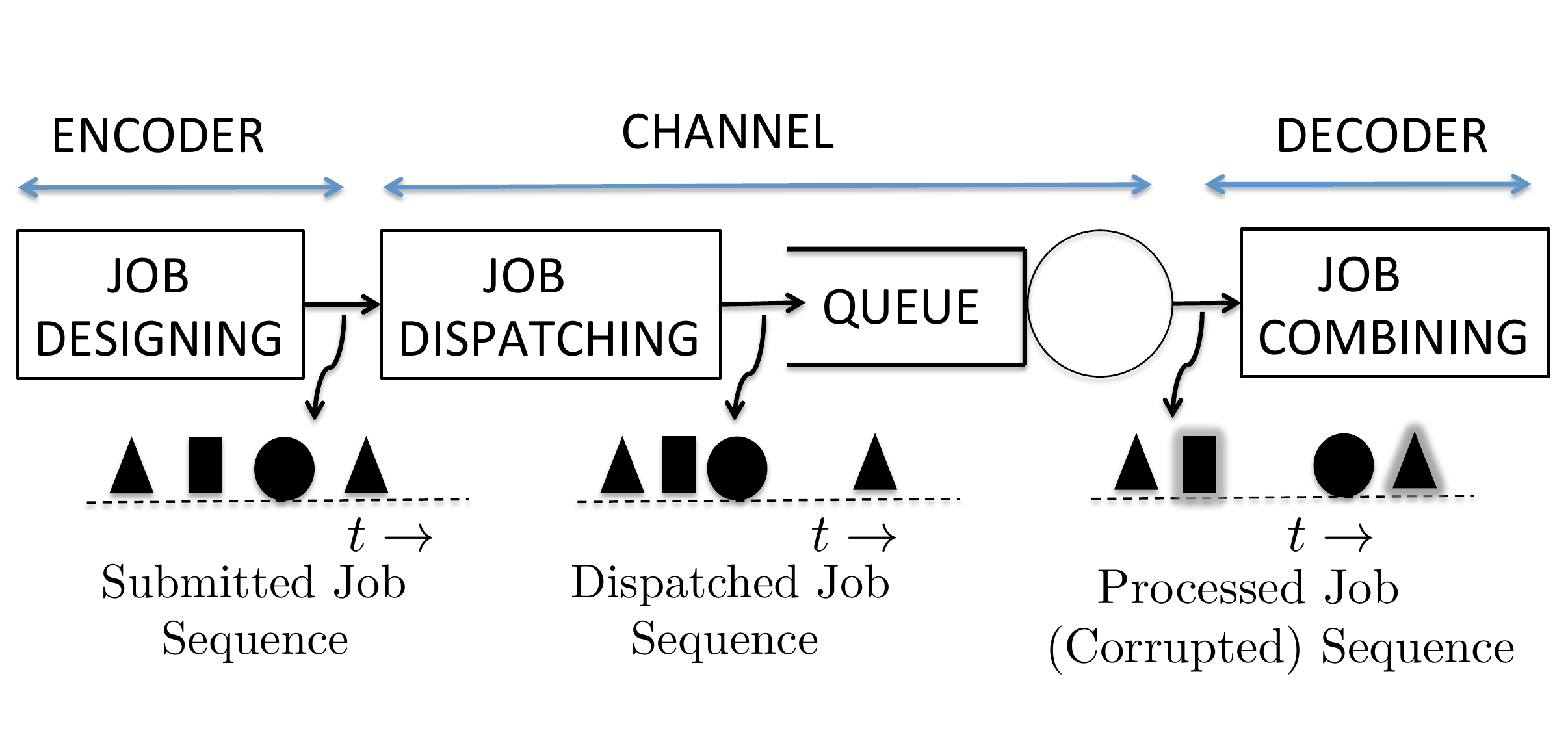}
  \caption{Schematic of the system.}
  \label{fig:schematic}
\end{figure}

Fig.~\ref{fig:schematic} presents a schematic of the system under study. As shown, there is an equivalence between our system and a communication channel. A large job is equivalent to a message in the communication setting. This large job is broken into a group of jobs, which is equivalent to a codeword of symbols. The random errors made by the server are equivalent to channel noise corrupting codeword symbols. Finally, the processed jobs (likely erroneous) are combined to complete the large job reliably, equivalent to decoding the original message from a received noisy codeword. Note that in contrast to many treatments of queuing systems, we are concerned with forward error correction rather than feedback-based repeat requests.  We study the limiting rate at which large jobs can be processed with arbitrarily small error probability,
and try to understand the best level at which to load the servers. 

This work lies at the intersection of information theory and queuing theory \cite{EphremidesH1998}, bringing together notions of burstiness and unreliability. Indeed, incorporating the information-theoretic notion of reliable job processing in a queueing system is of general theoretical interest. This problem also arises in many practical scenarios of growing prominence.

\begin{itemize}
\item {\bf Crowdsourcing and human computation:} Unlike machines or computers, the quality of service delivered by a human worker depends on his/her workload. Overloading a person with work often negatively impacts their quality of work \cite{Schwartz1978}, as in our model. Studying such scenarios is vital to understand optimal allocation of jobs in crowdsourcing.  In this context, the encoder in Fig.~\ref{fig:schematic} is the organization (e.g., Visipedia \cite{BransonVWPB2014}) that submits jobs to a crowdsourcing platform (e.g., Samasource \cite{BorokhovichCRVV2015}). The platform is the dispatcher and the crowd worker to whom jobs are assigned is the server.  Note that the transmitter and receiver are the same, and the code symbols represent \emph{completed work}. Error-correcting codes can be developed for difficult human computations, as described in \cite{VempatyVV2014}.

\item {\bf Multimedia communication:} When a user is in a live video or VoIP call over a multiple-access network, the access point---e.g., WiFi router or base station---has to contend for wireless resources to send the information packets. This results in an accumulation of packets at the MAC buffer of the access point. When the buffer is close to overflow, the access point either drops them \cite{SriramL1989}, sends their corresponding low-quality versions \cite{DraperTW2005} (assuming multiresolution coding \cite{Goyal2001b}), or packs multiple MAC packets in the available time slot using higher coding/modulation. All of these scenarios can be modeled by queue-dependent service quality.  We are interested in the maximum rate for reliable data transmission in this system.  Here the application provider (e.g.\ Google Voice or Skype) is the encoder, the core network of the internet is the dispatcher, and the wireless access node is the server. 

\item {\bf Stream computing:}  A queue-length based service quality model is also suitable for large-scale online learning with limited memory. In these settings often only a sketch of the data can be stored. As more data arrives, the quality of the sketches has to be reduced due to the memory constraint. This leads to a tradeoff between service quality and load on the system.  With emerging in-sensor computing devices, some basic error-control coding operations may be implemented before the data network dispatches information to the final server that performs appropriate computations.
\end{itemize} 

\subsection{Related Literature}
Anantharam and Verd\'{u} introduced the notion of timing channels for servers or queues \cite{AnantharamV1996}. Information is encoded in the times between consecutive information packets, and these packets are subsequently processed by a server according to some queueing discipline. Due to randomness in the sojourn times
of packets through servers, the encoded timing information is distorted, which the receiver must decode.  Later, a discrete-time version of the problem was studied \cite{BedekarA1998}; further insight into this problem was obtained by studying the entropy of arrival and departure processes of a queue \cite{PrabhakarG2003}.
Timing channels in queues have further been investigated for suitable decoding schemes \cite{SundaresanV2000}, zero-rate reliability \cite{WagnerA2005}, connections to game-theoretic settings \cite{GilesH2002}, information leakage \cite{GongKV2011,GorantlaKKCMK2012}, and models of information overload in microblogging \cite{TavanYB2013}.
Although we use some related proof techniques as these works, we are not concerned with information encoded in the timing between packets, only in the information in the symbols.
 
The study of information-theoretic limits of queuing multiple-access channels was pioneered by Telatar \cite{Telatar1992}, and further explored in \cite{TelatarG1995,RajTT2004}.  This line of work is essentially concerned with the reliable transmission of bursty sources \cite{MusyT2006}, as we are here.  In multiple-access settings, however, the main constraint beyond noise is interference among users.  The present work has a single user, but performance does degrade with greater burstiness,
a form of self-interference as it were.  A recent study of microbial communication also had a kind of self-interference called channel clogging \cite{MichelusiBEMM2015_arXiv}.

Queue-length dependent service times have been studied in operations research, see e.g.~\cite{Harris1967} and references thereto, but queue-length dependent service quality that we investigate here has remained unstudied.  In fact traditional queuing theory does not deal with issues of noise and quality, giving the present work theoretical novelty that may be applicable broadly in engineering theory. 

\subsection{Organization}
The remainder of the paper is organized as follows.  Sec.~\ref{sec:model} formalizes the system model, for two kinds of arrival processes we call Type I and Type II.  Sec.~\ref{sec:queueCapacity} defines the operational notion of system capacity and proves coding theorems that give an equivalent informational definition.  Sec.~\ref{sec:typeI} studies Type I capacities of well-known queueing systems and their relations with arrival and service processes, whereas Sec.~\ref{sec:typeII} studies Type II capacities.  Before the paper concludes, Sec.~\ref{sec:noTiming} considers the extension where timestamps are not available.

\section{System Model}
\label{sec:model}

A transmitter and a receiver {\em a priori} agree on a set of possible sequences of symbols (or codebook). The transmitter sends a sequence of symbols corresponding to a message to the dispatcher. The dispatcher sends these symbols to a server according to some stochastic process. The server services these symbols which are then received by the receiver. The receiver then tries to decode the message based on the received symbols.

The server works like a single first in first out (FIFO) queue with i.i.d.\ service requirements for each job. Jobs correspond to symbols from a finite field $\mathbb{F}$. In the model, servicing a job involves reading the symbol and outputting it. The server may make random errors during these steps and send out erroneous symbols. 

We are interested in the information capacity of such a system which we refer to as a \emph{queue-channel}.

\subsection{Queuing Discipline}

We consider a discrete-time system, $t \in \{0, 1, 2, \ldots\}$. Service requirement for jobs are i.i.d.\ and strictly positive, i.e.\, with values in $\mathbb{Z}_+$. The service time of the $i$th job is denoted $S_i$ and has a distribution $p_S$.

We use the following convention.
Arrivals at time $t$, if any, happen at the beginning of time slot $t$. Departures from the queue at time slot $t$, if any, happen at the end of the time slot. This implies that a job arriving at time slot $t$ may receive and possibly finish its service at time $t$.

Let $Q(t)$ be the number of jobs in the queue at the end of time slot $t$ and $Q_i$ be the number of jobs in the system when the $i$th job departs. As $S_i \ge 1$ for all $i$, at a time slot $t$, at most one job can depart.

We consider two basic types of arrivals processes (also called dispatch processes) into the queue: Type I and Type II. In a Type I process, there is at most one arrival in any time slot and the times between two consecutive arrivals are i.i.d.\ with distribution $p_A$ on $\{1, 2, \ldots \}$. In a Type II process, the numbers of arrivals $A(t)$ in time slot $t\ge 1$ are i.i.d.\ with distribution $m_A$ on $\{0, 1, 2, \ldots\}$. The service rate and arrival rate are $\mu$ and $\lambda$, respectively, satisfying $\EX_{p_S}[S] = 1/\mu$ and $\EX_{p_A}[A]= 1/\lambda$ or $\EX_{m_A}[A]=\lambda$, respectively. We assume $p_S$, $p_A$, and $m_A$ have finite second moments. 
For Type I systems, we assume either $p_A$ or $p_S$ has a support that spans $\Z_+$. For Type II systems, we assume $m_A(1)>0$.

\subsection{Service Noise}
Transmission of symbols from a finite field $\mathbb{F}$ over the queue-channel happens in two stages. Mapping the message, the transmitter sends symbols $\{X_i \in \mathbb{F}:1 \le i \le n\}$ to a dispatcher, which in turn sends the symbols to the server according to a stochastic process of arrival rate $\lambda$. For stability of the queue, we assume $\lambda<\mu$. 

The symbol corresponding to the $i$th symbol is $X_i \in \mathbb{F}$, 
and the output symbol corresponding to the $i$th symbol is 
$Y_i \in \mathbb{F}$.  They are related through the additive noise variable $Z_i \in \mathbb{F}$
representing work error, such that $Y_i = X_i + Z_i$.  The distribution of the errors $\{Z_i\}$ depends on $\{Q_i\}$. 
For any $i$, given $Q_i$, $Z_i$ is independent of any other processes or variables, and has a distribution $\psi_q$ (on $\mathbb{F}$) for $Q_i=q$.

An $n$-length transmission over the queue-channel is denoted as follows. 
Inputs are $\{X_i: 1 \le i \le n\}$, channel realizations are $\{Z_i: 1 \le i \le n\}$, and outputs are $\{Y_i: 1 \le i \le n\}$.
Throughout, a $k$-dimensional random vector is denoted by $U^k = (U_1, U_2, \ldots, U_k)$.

All logarithms in the paper have base $2$ so that information is measured in bits.

\section{Capacity of Queue-channel}
\label{sec:queueCapacity}
We are interested in the information capacity of unreliable server systems, i.e.\ the queue-channel described above. In this section, we present results that are generic, i.e.\ are true for both Type I and II arrivals.

\subsection{Definition}
Let $M, \hat{M} \in \mathcal{M}$ be the message to be transmitted and decoded, respectively. 
\begin{definition}
An $(n, \widetilde{R}, T)$ code consists of the encoding function $X^n=f(M)$ and the decoding function $\hat{M} = g(X^n, A^n, D^n)$, where the cardinality of the message set $|\mathcal{M}| = 2^{n\widetilde{R}}$, and for each codeword, the expected total time for all symbols to reach the receiver is less than $T$. 
\end{definition}
\begin{definition}
If the decoder chooses $\hat{M}$ with average probability of error less than $\epsilon$, that code is said to be $\epsilon$-achievable.
For any $0 < \epsilon < 1$, if there exists an $\epsilon$-achievable code $(n, \widetilde{R}, T)$, the rate $R = \frac{\widetilde{R}}{T}$ is said to be achievable.
\end{definition}
\begin{definition}
For an arrival process with distribution $p_A$ (Type I) or $m_A$ (Type II), the information capacity of the queue-channel is defined as the supremum over all achievable rates, which is denoted by $C(p_A)$ or $C(m_A)$ in bits per unit time.
\end{definition}
Since the transmitter sends symbols to the dispatcher first,
we assume the transmitter knows the arrival process statistics,
but not the realizations. Contrarily, the receiver knows the
realized arrival and departure times of each job.

\subsection{Coding Theorem}
\label{sec:codingthm}
Let $A_i \in \{1, \ldots\}$ and $D_i \in \{1, \ldots\}$ be the time of arrival into the queue and the time of departure from the queue of the $i$th symbol. The transmitter does not observe $\{A_i, D_i\}$, whereas the receiver observes these. Thus the queue-channel has inputs $\{X_i\}$ and outputs $\{Y_i, A_i, D_i\}$. As dispatch
is independent of job-design, the channel transition probability
factors as 
\[
\PR(Y^n, A^n, D^n|X^n) = \PR(A^n, D^n) \PR(Y^n|X^n, A^n, D^n)\mbox{.}
\]

The transmitter chooses $\{X_i\}$ and hence, can choose any joint distribution for the codebook described by $\{X_i\}$. Note $\{Y_i, A_i, D_i\}$ depends on $\{X_i\}$, as well as on the arrival and service processes. In general, $\{Y_i, A_i, D_i\}$ may not be a stationary process. This means that the queue-channel is not necessarily an information-stable channel \cite{Pinsker1964}, but the capacity formula can nevertheless be found using the information spectrum approach \cite{VerduH1994, Han2003}. Let the information density be $i(\cdot)$, the normalized information density be $$\frac{1}{n} i(X^n; Y^n, A^n, D^n) = \frac{1}{n} \log \frac{\PR(Y^n, A^n, D^n|X^n)}{\PR(Y^n, A^n, D^n)},$$ and the inf-information rate $\underline{\mathbf{I}} (\mbf{X}; \mbf{Y}, \mbf{A}, \mbf{D})$ be the \textit{lim-inf in probability} of the normalized information density, i.e.\, the largest $\alpha \in \mathbb{R} \cup \{\pm \infty\}$ such that for all $\epsilon > 0$, 
\begin{equation*}
\lim_{n \to \infty} \PR \left[ \frac{1}{n} i(X^n; Y^n, A^n, D^n) \le \alpha - \epsilon \right] = 0.
\end{equation*}

Then, capacity in bits per unit time of the queue-channel is given by
\begin{equation}
\label{eq:cap_exp1}
C(p_A) \mbox{ (and $C(m_A)$)} = \lambda \sup_{\PR(\mathbf{X})}\underline{\mathbf{I}} (\mbf{X}; \mbf{Y}, \mbf{A}, \mbf{D})\mbox{,}
\end{equation}
where $\lambda$ is the arrival rate defined in Sec.~\ref{sec:model}, and the supremum is over all input processes $\mathbf{X} = (X_1,X_2,\ldots)$.

This capacity expression is not easy to handle due to the various possibilities of $(A^n, D^n)$ that can arise, however, the next proposition allows us to characterize the distribution of $i(\cdot)$ (and hence, $\underline{\mathbf{I}}$) in a simpler form, in terms of the distributions of $X^n, Y^n$, and $Q^n$.

\begin{proposition}
\label{lem:simple_expression}
The capacity expression (\ref{eq:cap_exp1}) can be represented by using $Q^n$,
\begin{align*}
C(p_A) \mbox{ (and $C(m_A)$)} = \lambda \sup_{\PR(\mathbf{X})}\underline{\mathbf{I}} (\mbf{X}; \mbf{Y} | \mbf{Q}). 
\end{align*}
\end{proposition} 
\begin{IEEEproof}
It suffices to show that 
\begin{align}
 i(X^n; Y^n, A^n, D^n) = i(X^n; Y^n | Q^n). \label{eq:ADQ1}
\end{align}
 Note that the additive noise $Z^n$ depends only on $Q^n = \phi_n(A^n,D^n)$, where $\phi_n(\cdot)$ is a function that computes the number of symbols in the queue. Hence, $\PR(Y^n|A^n,D^n,X^n) = \PR(Y^n|Q^n,X^n)$. Also, $X^n$ is independent of $(A^n,D^n)$.
\begin{align*}
&\frac{\PR(Y^n,A^n,D^n|X^n)}{\PR(Y^n,A^n,D^n)} = \frac{\PR(A^n,D^n|X^n) \PR(Y^n|A^n,D^n,X^n) }{\PR(A^n,D^n)\PR(Y^n|A^n,D^n)} \\
&= \frac{\PR(Y^n|A^n,D^n,X^n) }{\PR(Y^n|A^n,D^n)} = \frac{\PR(Y^n|Q^n,X^n) }{\PR(Y^n|A^n,D^n)} \\
&= \frac{\PR(Y^n|Q^n,X^n) }{\sum_{X^n} \PR(Y^n,X^n|A^n,D^n)} = \frac{\PR(Y^n|Q^n,X^n) }{\sum_{X^n} \PR(X^n|A^n,D^n)\PR(Y^n|A^n,D^n,X^n)} \\
&= \frac{\PR(Y^n|Q^n,X^n) }{\sum_{X^n} \PR(X^n|Q^n)\PR(Y^n|Q^n,X^n)} = \frac{\PR(Y^n|Q^n,X^n) }{\PR(Y^n|Q^n)}.
\end{align*}
Taking logarithm and normalizing yields $i(X^n; Y^n, A^n, D^n) = i(X^n; Y^n | Q^n)$.
\end{IEEEproof}
Thus, it follows that the distribution of $i(\cdot)$ depends only on the joint distribution of $(X^n, Y^n, Q^n)$. 

Based on this, we can give a single-letter characterization of the capacity of the queue-channel.  In the proof of the forthcoming coding theorem, the converse part is essentially due to Fano's inequality and basic properties of information quantities \cite{CoverT1991}. The direct part follows by choosing an appropriate input processes $\mathbf{X}$ to lower bound $\sup_{\PR(\mathbf{X})} \underline{\mathbf{I}} (\mbf{X}; \mbf{Y} | \mbf{Q})$. In this regard, this proof is structurally similar to earlier work that applied information spectrum techniques, e.g.~\cite{CaireS1999, BedekarA1998}.

The proof of the coding theorem also implicitly depends on the following lemma which characterizes the process $\{Q_i\}$. 
\begin{lemma}
\label{lem:Qstationary}
Under the assumptions in Sec. \ref{sec:model} and $\lambda<\mu<1$, there exists a unique distribution $\pi$ such that if $Q_1 \sim \pi$, then $Q_i \sim \pi$ for all $i \ge 1$, and the process $\{Q_i\}$ is ergodic, i.e.\, for any $f:\{0,1,\ldots\} \to \mathcal{R}$ with finite $\EX_\pi f$, almost surely $\frac{1}{n} \sum_{i=1}^n f(Q_i) \to \EX_\pi f$ as $n \to \infty$. Moreover, for any initial distribution of $Q_1$, $Q_i$ converges to $\pi$ in distribution and $\pi(q)>0$ for all $q \in \{0,1,\ldots\}$.
\end{lemma} 
\begin{IEEEproof}
See Appendix~\ref{app:lem_qstationary}.
\end{IEEEproof}

Now the capacity theorem.
\begin{theorem}
\label{thm:queueCapacity}
For a given arrival process distribution $p_A$ (or $m_A$) with $\lambda < \mu < 1$  which follows the assumption in Sec. \ref{sec:model}, there exists a distribution $\pi$ such that $\pi(q)>0$ for all $q \in \{0,1,\ldots\}$ and $\PR(Q_n) \to \pi$ as $n\to \infty$. The capacity of this queue-channel is $\lambda(\log |\mathbb{F}| - \sum_q \pi(q) H(\psi_q))$, where $H(\psi_q)$ is the entropy of a distribution $\psi_q(Z)$ on any finite set of size $|\mathbb{F}|$.
\end{theorem}
\begin{IEEEproof}
First, we prove the converse. Let $\tilde{R}$ be the rate in bits per symbol. Let $M$ and $\hat{M}$ be transmitted and decoded messages, respectively, such that $\hat{M} = g(Y^n, A^n, D^n)$, where $g(\cdot)$ is a decoding function. Note that a Markov chain $M-X^n-(Y^n, A^n, D^n)-\hat{M}$ holds.

\begin{align}
n \tilde{R} &= H(M) = I(M;\hat{M}) + H(M|\hat{M}) \nonumber \\
& \le I(M;\hat{M}) + n \epsilon \nonumber \\
& \le I(X^n;Y^n,A^n, D^n) + n \epsilon \nonumber \\
& = I(X^n;Y^n|Q^n) + n \epsilon \label{eq:ADQ2} \\
& = H(Y^n|Q^n) - H(Y^n|X^n, Q^n) + n \epsilon \nonumber \\
& = \sum_{i=1}^n H(Y_i|Q^n, Y^{i-1}) - \sum_{i=1}^n H(Y_i|X^n, Q^n, Y^{i-1}) + n \epsilon \nonumber \\
& \le \sum_{i=1}^n H(Y_i|Q_i) - \sum_{i=1}^n H(Y_i|X_i, Q_i) + n \epsilon. \label{eq:converse1}
\end{align}
The first two inequalities are Fano's inequality and the data processing inequality. $I(X^n;Q^n)=0$ since $X^n$ is chosen independently of the dispatch and departure processes $(A^n,D^n)$. Eq. \eqref{eq:ADQ2} follows since $I(X^n;Y^n,A^n,D^n)=\EX[n  i(X^n;Y^n,A^n,D^n)]$ and \eqref{eq:ADQ1}. The last inequality follows since removing conditioning only increases entropy and $Y_i$ depends only on $X_i, Q_i$.

Note that 
\begin{align*}
H(Y_i|q) - H(Y_i|X_i,q) & \le \log|\mathbb{F}| - H(\psi_q) \mbox{,}
\end{align*}
with equality attained by assuming $X_i$ is uniformly distributed over $\mathbb{F}$. Let $p_i$ be the distribution of the queue-length seen by the $i$th departure. Then, \eqref{eq:converse1} can be further upper bounded as:
\begin{align*}
n \tilde{R} &\leq \sum_{i=1}^n H(Y_i|Q_i) - \sum_{i=1}^n H(Y_i|X_i, Q_i) + n \epsilon \\
&= \sum_{i=1}^n \left( \sum_q p_i(q) (H(Y_i|q) - H(Y_i|X_i,q)) \right) + n \epsilon \\
&\le n \left( \log|\mathbb{F}| - \frac{1}{n} \sum_{i=1}^n \sum_q p_i(q) H(\psi_q) \right) + n \epsilon \\
&= n \log|\mathbb{F}| - n \sum_q \pi_q H(\psi_q) + \delta_n + n \epsilon,
\end{align*}
where $\delta_n \to 0$ as $n \to \infty$. The last step follows because the queue-length process is ergodic (Lemma \ref{lem:Qstationary}), and hence, the Cesaro mean of expectations at different times converges to the expectation with respect to the stationary distribution. Finally multiplying by $\lambda$ concludes the converse.

To show achievability, we pick $X^n$ i.i.d.\ uniformly at random from $\mathbb{F}$ and show that the inf-information rate in Proposition \ref{lem:simple_expression} for this process is $\log|\mathbb{F}| - \sum_q \pi_q H(\psi_q)$.

Note that as $\mathbb{F}$ is a field, for any element $y_i \in \mathbb{F}$, $y_i-X_i$ spans all elements in $\mathbb{F}$. Hence, $\sum_{X \in \mathbb{F}} \psi_{Q_i}(Y_i-X_i) = 1$. Thus,
\begin{align*}
\PR(Y_i|Q_i) &= \sum_{X_i \in \mathbb{F}} \PR(Y_i, X_i|Q_i) = \sum_{X_i \in \mathbb{F}} \PR(X_i|Q_i) \PR(Y_i|X_i,Q_i) = \sum_{X_i \in \mathbb{F}} \frac{1}{|\mathbb{F}|} \psi_{Q_i}(Y_i-X_i) = \frac{1}{|\mathbb{F}|}.
\end{align*}
Since the $\{X_i\}$ are i.i.d.\ and given $Q_i$, $Y_i$ depends only on $X_i$, we get the following product form: $\PR(y^n|x^n, q^n)=\prod_i \PR(y_i|x_i, q_i)$, $\PR(y^n|q^n)=\prod_i \PR(y_i|q_i)$.  This implies:
\begin{align*}
\frac{1}{n} i(X^n; Y^n|Q^n) &= \frac{1}{n} \log \frac{\PR(Y^n|X^n, Q^n)}{\PR(Y^n|Q^n)} = \frac{1}{n} \log \frac{\prod_i \PR(Y_i|X_i, Q_i)}{\prod_i \PR(Y_i|Q_i)} = \frac{1}{n} \log \frac{\prod_i \psi_{Q_i}(Z_i) }{\prod_i 1/|\mathbb{F}|} \\
&= \log |\mathbb{F}| + \frac{1}{n} \sum_{i=1}^n \log \psi_{Q_i}(Z_i)\mbox{.}
\end{align*}
Now we use Lemma~\ref{lem:Qstationary} to take a limit.
\begin{align*}
\frac{1}{n} i(X^n; Y^n|Q^n) &\to \log |\mathbb{F}| + \EX_{\pi_Q,Z}[\log \psi_{Q}(Z)] \mbox{ almost surely as } n \to \infty \\
&= \log|\mathbb{F}| - \sum_q \pi_q H(\psi_q)\mbox{.}
\end{align*}
\end{IEEEproof}

With the coding theorems developed in this section in hand, Secs.~\ref{sec:typeI} and \ref{sec:typeII} study the capacity of a few interesting classes of discrete-time queues. This results in insights regarding the dispatch and service processes that result in best and worst information processing rates. 

\subsection{Comments}
\begin{figure}
	\centering
	\includegraphics[scale=0.7]{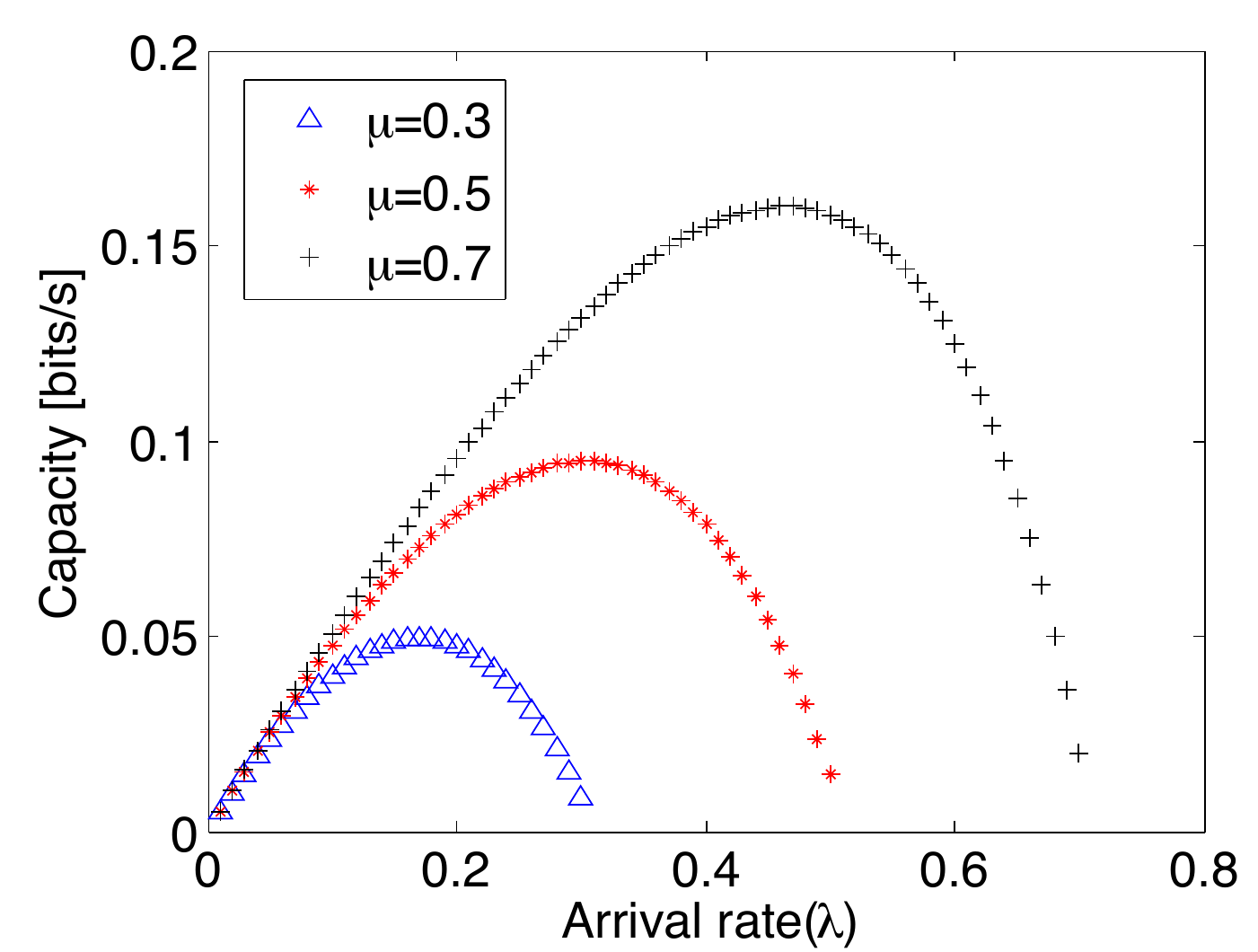}
	\caption{Capacity of $\geo/\geo/1$ queue is plotted against arrival rate (for different service rates) for $\mathbb{F}=\{0,1\}$ and noise distribution $\PR(Z=1)=0.1$ for $q=0$, otherwise $\PR(Z=1)=0.4$.}
	\label{fig:performance}
\end{figure}

Before studying specific classes of queuing systems, we comment on the relation between the maximum packet throughput and the maximum information throughput (the notion of capacity defined here) of a queuing system. Packet throughput of a queuing system is the maximum rate of packet arrivals that can be served without instability; hence the packet throughput increases with $\lambda$ on $[0,\mu)$. Though the expression for capacity (information throughput) has $\lambda$ as a multiplicative factor, this does not mean that information throughput increases with $\lambda$. In typical queuing systems, the survival function corresponding to the stationary probability is increasing in $\lambda$. Thus, an increase in $\lambda$ also has a negative impact on the terms involving $\pi$. Hence, in typical queuing systems, there is an optimal $\lambda \in (0,\mu)$ that maximizes information throughput. Fig.~\ref{fig:performance} shows an example.

\section{Queues with Type I arrival}
\label{sec:typeI}
This section is devoted to understanding the capacity of a queue with a Type I arrival process and its dependence on the distribution of service times and inter-arrival times. First, we find the capacity of a queue with geometric service time and arbitrary arrival process, and characterize the capacity-optimizing arrival distributions. Then, we study the capacity of a queue with geometric inter-arrival time and find the capacity-optimizing service time distribution. Capacity has
a saddle point behavior around the geometric distribution.

In the application scenarios discussed in Sec.~\ref{sec:intro}, server performance deteriorates with increasing queue-length. 
Deterioration of server performance with increasing queue-length is captured by a $\{\psi_q\}$ whose entropy is non-decreasing with $q$. A $\{\psi_q\}$ of practical interest is a threshold behavior of the error-entropy with increasing queue-length: $H(\psi_q)=h_0$ for $q \le b$ and $H(\psi_q)=h_{b+1}$ for $q \ge b+1$, for some $b \in \{0,1,\ldots\}$. 

Threshold behavior captures a state of server
panic based on workload, suitable for human servers and wireless access points with small MAC buffer. The special case of $b=0$ describes a human server that is distracted by any waiting job or a bufferless MAC. The special case of $b=1$ corresponds to a human server being
distracted if more than one job is waiting.

\subsection{Discrete-time $G/\geo/1$ queue}
\label{subsec:GIGeo1}
For a $G/\geo/1$ queue, the service time distribution is geometric with an expected service time $\frac{1}{\mu}$, $\mu < 1$. The arrival process is Type I, with the inter-arrival times distributed as $p_A$ and the expected time between arrivals $\frac{1}{\lambda}$, $\lambda < \mu$.  Since this queueing system satisfies the assumptions in Sec.~\ref{sec:model}, its capacity can be obtained from Theorem \ref{thm:queueCapacity}. For any arrival distribution, the capacity of $G/\geo/1$ queue is given by the following theorem.

\begin{theorem}
\label{thm:GGeo1}
The capacity of the $G/\geo/1$ queue-channel is 
$\lambda(\log|\mathbb{F}| - (1-\sigma) \sum_q \sigma^q H(\psi_q))$, where $\sigma$ is the unique solution of the equation $x = \sum_{n=0}^\infty p_A(n) (1-\mu+x\mu)^n$ in $(0,1)$.
\end{theorem}
\begin{IEEEproof}
See Appendix \ref{app:pf_thm2}.
\end{IEEEproof}
Proof of this theorem involves obtaining the steady-state distribution $\pi$ of the queue-lengths seen by the departures.
Towards this, techniques similar to that in the analysis of continuous-time $GI/M/1$ queues \cite{Kleinrock1975_I} are extended to the discrete-time setting. The closed-form expressions here differ to some extent from that in $GI/M/1$. Also, note that some of the intermediate steps in the proof of Theorem \ref{thm:GGeo1} are used to prove some later results. 

Based on the capacity characterization of the $G/\geo/1$ queue, we explore the space of arrival distributions. This leads to the following result about the best and worst (in terms of capacity) arrival distribution for a $G/\geo/1$ queue.

\begin{proposition}
\label{prop:GGeo1Min} 
For $G/\geo/1$ queue with thresholded noise such that $H(\psi_0)=\cdots=H(\psi_b)<H(\psi_{b+1})=\cdots$ for some $b\in \{0,1,\ldots\}$, deterministic inter-arrival time maximizes capacity among all arrival distributions with the same $\lambda$, for $\frac{1}{\lambda} \in \Z_+$.
\end{proposition}

\begin{IEEEproof}
Proof of this result builds on the property of the fixed point equation $x=\sum_{t=0}^\infty (1-\mu+\mu x)^t p_A(t)$, and uses an intermediate result in the proof of Theorem \ref{thm:GGeo1}.

First see that for any arrival distribution $p_A$, $\pi(q) = (1-\sigma) \sigma^q$, and capacity is $\log|\mathbb{F}| - \sum_q \pi(q) H(\psi_q)$, which is maximized when $(1-\sigma) \sum_q \sigma^q H(\psi_q)$ is minimized. Since noise is thresholded at $b$, i.e.\, $h_0 = H(\psi_0) = \cdots = H(\psi_b)$ and $h_{b+1} = H(\psi_{b+1})= \cdots$, then the latter term may be written as 
\begin{align*}
(1-\sigma) \sum_q \sigma^q H(\psi_q) = h_0 (1-\sigma^{b+1}) + h_{b+1}(1- (1-\sigma^{b+1})) = h_0 + (h_{b+1}-h_0)\sigma^{b+1}.
\end{align*}
Hence, for a given $\{\psi_q\}$, capacity is maximized when $\sigma$ is minimized.

Next, note that the curves $\tilde{A}(\sigma) =\sum_{t=1}^\infty p_A(t) (1-\mu+\mu\sigma)^t$ are convex and increasing with $\sigma$, and $\tilde{A}(0)>0$ (see Lemma \ref{lem:xConvex} in Appendix). Also, there is a unique fixed point in $(0,1)$. Thus, for these classes of curves, the curve that lower bounds a set of curves crosses the line $y=\sigma$ at the smallest value of $\sigma$ among that set of curves. Similarly, the curve that upper bounds a set of curves crosses the line $y=\sigma$ at the largest value of $\sigma$.

For any $0<\alpha<1$ and any distribution $p_A$ with mean $\frac{1}{\lambda}$,
$$\sum_{t=0}^\infty \alpha^t p_A(t) \ge \alpha^{\frac{1}{\lambda}}\mbox{,}$$ by Jensen's inequality, as $\alpha^t$ is convex. Thus for any $\sigma \in (0,1)$ and $p_A$ with mean $\frac{1}{\lambda}$,
\begin{align*}
\tilde{A}(p_A,\sigma) &= \sum_{t=0}^\infty (1-\mu+\mu \sigma)^t p_A(t) \\
& \ge (1-\mu + \mu \sigma)^{\frac{1}{\lambda}} \\
&= \tilde{A}(\mathsf{det},\sigma),
\end{align*}
where the equality can be attained by a deterministic inter-arrival time. This implies that the curve $\tilde{A}(\mathsf{det},\sigma)$ is a lower-bounding curve for all other curves corresponding to different $p_A$. 
\end{IEEEproof}

\begin{proposition}
\label{prop:global_min}
For the $G/\geo/1$ queue with $\{\psi_q\}$ such that $H(\psi_0)=\cdots=H(\psi_b)<H(\psi_{b+1})=\cdots$ for some $b\in \{0,1,\ldots\}$, $\tilde{p}_A(t, \epsilon)$ asymptotically minimizes the capacity among all arrival processes as $\epsilon \to 0$, where 
\begin{align*}
\tilde{p}_A(t, \epsilon) = \begin{cases} 1-\epsilon, & t=1 \\ \epsilon, & t=N(\epsilon), \end{cases}
\end{align*}
for $\epsilon > 0$ and $N(\epsilon)$ is chosen to satisfy the mean constraint $1/\lambda$.

\end{proposition}
\begin{IEEEproof}
It is sufficient to show that $\tilde{A}(p_A,\sigma)$ is asymptotically maximized by $\tilde{p}_A(t,\epsilon)$ as $\epsilon \to 0$.

Consider developing an upper bound of $\tilde{A}(p_A,\sigma)$ first. Using the fact that for $\alpha \in (0,1), \alpha^t$ is decreasing,
\begin{align*}
\tilde{A}(p_A,\sigma) &= \sum_{t=0}^\infty (1-\mu+\mu \sigma)^t p_A(t) \le \sum_{t=0}^\infty (1-\mu+\mu \sigma) p_A(t) = (1-\mu+\mu \sigma).
\end{align*}

On the other hand, $\tilde{A}(p_A,\sigma)$ evaluated at $\tilde{p}_A(t,\epsilon)$ is:
\begin{align*}
\tilde{A}(\tilde{p}_A(t,\epsilon),\sigma) = (1-\mu+\mu \sigma)(1-\epsilon) + (1-\mu+\mu \sigma)^N \epsilon,
\end{align*}
which approaches the upper bound as $\epsilon \to 0$, but has a fixed-point solution in $(0,1)$. The pmf $\tilde{p}_A(t,\epsilon)$ asymptotically maximizes the fixed-point solution as $\epsilon \to 0$, thus minimizing the capacity.
\end{IEEEproof}

The results of Propositions \ref{cor:GeoG1Thre0} and \ref{prop:global_min} agree with our intuition. Deterministic arrivals in Proposition \ref{cor:GeoG1Thre0} give enough time to the server with a given service rate, so that each job sees the lowest queue length behind it on average. On the other hand, a typical realization of $\tilde{p}_A(t,\epsilon)$ is that jobs arrive every time slot (corresponding to $t=1$) for some time interval but then the next job arrives a very long time later corresponding to $t=N(\epsilon)$. The server will be busiest during the first interval, but will be almost idle until the next job. It yields the worst performance.

In crowdsourcing, it is common for the arrival process to come from some kind of job pre-processing. Since this pre-processing system itself could be serial or parallel chains of servers with exponentially-distributed random delays, we are interested in classes of arrival processes that are certain geometric families of distributions.

Let $\{A_{i}, 1 \le i \le I\}$ be independent geometric random variables with means $\frac{1}{\lambda_{i}}$. Then define $A^s$ to be a sum-of-geometric random variable and  to be $\mathcal{A}^s$ the set of such probability distributions with mean $\frac{1}{\lambda}$, i.e.\,
\begin{align*}
A^s &=\sum_i A_i \\
\mathcal{A}^s &= \left\{p_{A^s}: \EX [A^s]=\frac{1}{\lambda} \right\}.
\end{align*}
Also define $A^m$ to be a mixture of geometric random variables such that $A^m=A_i$ with probability mass $\{c_i\}$ whose support is $\{1 \le i \le I\}$, with $\mathcal{A}^m$ as the set of such probability distributions with mean $\frac{1}{\lambda}$, i.e.\,
\begin{align*}
A^m &=A_i \mbox{ with probability } c_i, \\
\mathcal{A}^m &= \left\{p_{A^m}: \EX [A^m]=\frac{1}{\lambda} \right\}.
\end{align*}
Then the next lemma follows.

\begin{lemma} \label{lem:geoProdSum}
For any $p_{A^s} \in \mathcal{A}^s$,
\begin{equation*}
\tilde{A}(p_{A^s}, \sigma) \le \tilde{A}(\geo, \sigma).
\end{equation*}

On the other hand, for any $p_{A^m} \in \mathcal{A}^m$.
\begin{equation*}
\tilde{A}(p_{A^m}, \sigma) \ge \tilde{A}(\geo, \sigma).
\end{equation*}
\end{lemma}

\begin{IEEEproof}
See Appendix \ref{app:pf_lem_prodsum}.
\end{IEEEproof}

\begin{proposition}
\label{prop:GGeo1Max}
For any $G/\geo/1$ queue-channel with thresholded noise at $b\in \{0,1,\ldots\}$, geometric inter-arrival times minimize and maximize capacity among all arrival distributions of $\mathcal{A}^s$ and $\mathcal{A}^m$, respectively.
\end{proposition}
\begin{IEEEproof}
Following the arguments in the proof of Proposition \ref{prop:GGeo1Min}, we only need to show that geometric inter-arrival times maximizes (resp.~minimize) $\sigma$ for a given $\lambda$ among $\mathcal{A}^s$ (resp.~$\mathcal{A}^m$). Then the proposition follows from Lemma \ref{lem:geoProdSum}.
\end{IEEEproof}

There is an important takeaway from this result in the context of job pre-processing for crowdsourcing. In crowdsourcing systems all jobs are pre-processed to make them suitable for crowd workers and the inter-arrival (inter-dispatch) time in our model corresponds to this pre-processing time. The above theorem implies it is best to have a deterministic pre-processing time. However, if pre-processing times are highly variable due to some system issues (geometric is the most entropic), then instead of having a single pre-processing step it is better to have a series of sub-steps (corresponding to sum-of-geometric) for pre-processing.

A corollary is the capacity extrema representation of the $G/\geo/1$ queue-channel among the class of sum- (resp.~mixture-) of-geometric distributions.
\begin{corollary}
\label{cor:maxCapacityGGeo1}
For a given arrival rate $\lambda$ and given service rate $\mu$, the minimum (resp.~maximum) capacity of $G/\geo/1$ queue-channel among the class of sum- (resp.~mixture-) of-geometric inter-arrival distributions is $\lambda(\log|\mathbb{F}| - (1-\sigma^*) \sum_q {\sigma^*}^q H(\psi_q))$, where $\sigma^* = \frac{\lambda(1-\mu)}{\mu(1-\lambda)}$.
\end{corollary}
\begin{IEEEproof}
From the proof of Proposition \ref{prop:GGeo1Max}, we know that geometric arrival achieves capacity extrema for arrival rate $\lambda$,
$$\sum_{t=0}^\infty \alpha^t p_A(t) = \frac{\frac{\alpha}{1-\alpha}}{\frac{1}{\lambda}+\frac{\alpha}{1-\alpha}}.$$
Letting $\alpha = 1-\mu+\mu\sigma$ and solving the fixed point equation
$$\tilde{A}(\geo, \sigma) =  \frac{\frac{1-\mu+\mu\sigma}{\mu-\mu\sigma}} {\frac{1}{\lambda} + \frac{1-\mu+\mu\sigma}{\mu-\mu\sigma}} = \sigma,$$ we have the unique solution 
$\sigma^* = \frac{\lambda(1-\mu)}{\mu(1-\lambda)}$.
\end{IEEEproof}

\subsection{Discrete-time $\geo/G/1$ queue}
\label{subsec:GeoG1}
In this section we consider another important class of queues, for which the arrival process is Bernoulli, i.e.\, inter-arrival times are geometric, but the service times have a general distribution.

Define ${n \choose k} = 0$ if $n < k$. By characterizing the stationary distribution seen by departures, we prove the following capacity result.
\begin{theorem}
\label{thm:GeoG1}
For $j \in \{0, 1, \ldots\}$, let $k_j = \sum_{t=0}^\infty {t \choose j} (1-\lambda)^{t-j} \lambda^j p_S(t)$, and for all complex $z$ with
$|z|<1$, $K(z)=\sum_{j=0}^\infty z^j k_j$, then capacity of this system is $\lambda(\log|\mathbb{F}| - \sum_q \pi_q H(\psi_q))$ for 
$\pi_0=1-\frac{\lambda}{\mu}$, and $\pi_k=\lim_{z \to 0}
\frac{\Pi(z) - \sum_{j=0}^{k-1}\pi_j z^j}{z^k}$, where $\Pi(z) = (1-\frac{\lambda}{\mu})\frac{(z-1)K(z)}{z-K(z)}$.
\end{theorem}
\begin{IEEEproof}
See Appendix \ref{app:pf_thm3}.
\end{IEEEproof} 
Derivation of the stationary distribution here follows similar steps as the derivation of the stationary distribution of $M/G/1$ queue \cite{Kleinrock1975_I}.

Next, we investigate the service time distributions that respectively maximize or minimize the capacity of a $\geo/G/1$ queue-channel. 
First, we consider the case of threshold error-entropy behavior for threshold $b=0$. The following corollary is a direct consequence of Theorem \ref{thm:GeoG1}.

\begin{corollary}
\label{cor:GeoG1Thre0}
If $\{\psi_q\}$ are such that $H(\psi_0) < H(\psi_1) = H(\psi_2) = \cdots$, then the capacity of the $\geo/G/1$ queue is the same for all $p_S$.
\end{corollary}
\begin{IEEEproof}
The capacity in this case depends on $\pi$ only through $\pi_0$, but $\pi_0$ is the same for all service distributions with mean $\frac{1}{\mu}$, as $\pi_0=1-\frac{\lambda}{\mu}$.
\end{IEEEproof}

For the case with threshold $b=1$, it can be shown that different service time distributions result in different capacities.

\begin{proposition}
\label{prop:GeoG1Thre1}
If $H(\psi_0) = H(\psi_1) < H(\psi_2) = H(\psi_3) = \cdots$, then the capacity of the $\geo/G/1$ queue is maximized by a deterministic service time (for $\frac{1}{\mu} \in \Z_+$) and is asymptotically minimized by $\tilde{p}_S(t, \epsilon)$ as $\epsilon \to 0$, where
\begin{align*}
\tilde{p}_S(t, \epsilon) = \begin{cases} 1-\epsilon, & t=1 \\ \epsilon, & t=N(\epsilon), \end{cases}
\end{align*}
for $\epsilon>0$ and $N(\epsilon)$ is chosen to satisfy the mean constraint $1/\mu$. Among the class of sum-of-geometric random variables, capacity is minimized by the geometric service time distribution.
\end{proposition} 
\begin{IEEEproof}
Let $h_0=H(\psi_0)$ and $h_2=H(\psi_2)$, $h_0<h_2$. Then, by Theorem \ref{thm:queueCapacity}, the capacity of the system is $\lambda(\log|\mathbb{F}| - h_0(\pi_0+\pi_1) - h_2(1-\pi_0-\pi_1))$. 

It is clear from the capacity expression, that it is maximized (resp.~minimized) when $\pi_0+\pi_1$ is maximized (resp.~minimized). Hence, it is enough to prove that deterministic service time maximizes $\pi_0+\pi_1$, and geometric service time minimizes $\pi_0+\pi_1$ among the class of sum-of-geometric random variables.

Note that
$$\pi_1 = \lim_{z \to 0} \frac{\Pi(z) - \pi_0}{z},$$
which, after a few steps of algebra using the expression for $\Pi(z)$ and the fact that $\pi_0=1-\frac{\lambda}{\mu}$, gives  
$$\pi_1 = (1-\tfrac{\lambda}{\mu}) \frac{1-K(0)}{K(0)}.$$
Thus, $\pi_0 + \pi_1 = (1-\tfrac{\lambda}{\mu}) \frac{1}{K(0)}$. Using the definition that $K(0)=k_0=\PR(\mbox{no arrivals in} \ S)$, the capacity is minimized when $k_0$ is maximized and vice-versa. After decomposing $k_0$ for all $t$,
$$k_0 = \sum_{t=1}^\infty (1-\lambda)^t p_S(t).$$

The conclusion follows from the proofs of Lemma \ref{lem:geoProdSum} and Proposition \ref{prop:global_min}: the deterministic arrival with mass at $\frac{1}{\mu} \in \mathbb{Z}_+$ minimizes $k_0$ by Jensen's inequality, and $\tilde{p}_S(t, \epsilon)$ asymptotically maximizes $k_0$ as $\epsilon \to 0$. In addition, $k_0$ is maximized by geometric distribution among the class of sum-of-geometric random variables by Lemma \ref{lem:geoProdSum}.
\end{IEEEproof}

Proposition \ref{prop:GeoG1Thre1} says that handling works with regularity in time yields the least queue length on average; cramming and staying idle is the worst.
For thresholded noise behavior we observe the following. For a geometric service time, the worst dispatch
process among the sum-of-geometric distributions is geometric. On the other hand, for a geometric arrival process,
the geometric service time is the worst among the sum-of-geometric distributions. Thus, if we visualize the capacity
function of a single server queue for a given arrival and service rate plotted against arrival and service distributions
(restricted to sum-of-geometric), there is a minimum where both distributions are geometric.

In the context of crowdsourcing, this means it is always better to split highly-variable pre-processing (corresponding
to the dispatch process) or human work (corresponding to the service process) steps into a series of sub-steps. That is, it is always better to take a job by parts (if coordination costs are not too high \cite{ChatterjeeVV2015}).

\section{Queues with Type II arrivals}
\label{sec:typeII}
In this section, we study the queue-capacity of systems with Type II arrivals.  
An equivalent capacity expression holds for Type II arrivals, i.e.\, possibly multiple arrivals in a time slot. Let $N_i$ be a random variable counting the number of arrivals at time $i$. Thus, the $\{N_i\}$ are i.i.d.\ with distribution $m_A$.  
\begin{theorem}
\label{thm:GxG1}
The queue-channel capacity of a queue with Type II arrivals distributed as $m_A$, and service time distributed as $p_S$ is given by 
$\lambda(\log|\mathbb{F}| - \sum_q \pi_q H(\psi_q))$, for 
$\pi_0=1-\frac{\lambda}{\mu}$, and $\pi_k=\lim_{z \to 0}
\frac{\Pi(z) - \sum_{j=0}^{k-1}\pi_j z^j}{z^k}$, where $\Pi(z) = (1-\frac{\lambda}{\mu})\frac{(z-1)K(z)}{z-K(z)}$, $k_j = \sum_{t=1}^\infty \PR(\sum_{i=1}^t {N_i}=j) p_S(t)$.
\end{theorem}
\begin{IEEEproof}
The probability of $j$ arrivals within a service time is $\sum_{t=1}^\infty \PR(\sum_{i=1}^t {N_i}=j) p_S(t)$. The remainder of the proof follows the same approach as the proof of Theorem \ref{thm:GeoG1}.
\end{IEEEproof}

\subsection{Effects of Service Processes}
First, we characterize the effect of different service processes on the capacity, for a given arrival process. As the following results show, deterministic service is best and bursty service is worst, as in Proposition \ref{prop:GeoG1Thre1}.
\begin{proposition}
\label{prop:GxG1Thre1}
Suppose that $H(\psi_0)=H(\psi_1)<H(\psi_2)=\cdots$. For a given Type II arrival process $m_A$, the maximum capacity is achieved by deterministic service time over all service time distributions. The minimum capacity is asymptotically achieved by $\tilde{p}_S(t,\epsilon)$ as $\epsilon \to \infty$, where
\begin{align*}
\tilde{p}_S(t, \epsilon) = \begin{cases} 1-\epsilon, & t=1 \\ \epsilon, & t=N(\epsilon), \end{cases}
\end{align*}
for $\epsilon > 0$ and $N(\epsilon)$ is chosen to satisfy the mean constraint $1/\mu$.
In addition, the minimum capacity among the class of sum-of-geometric random variables is achieved by geometric service time distribution.
\end{proposition}
\begin{IEEEproof}
Following the proof of Proposition \ref{prop:GeoG1Thre1}, we only need to prove that $k_0$ is minimized and asymptotically maximized by deterministic service time and $\tilde{p}_S(t, \epsilon)$, respectively. Further, $k_0$ needs to be maximized by geometric service time among the class of sum-of-geometric random variables.

Note that $k_0=\sum_{t=1}^\infty \PR(\sum_{i=1}^t {N_i}=0) p_S(t) = \sum_{t=1}^\infty (m_A(0))^t p_S(t)$, where $0<m_A(0)<1$. Hence, the results follow from the proof of Proposition \ref{prop:GeoG1Thre1}.
\end{IEEEproof}

\subsection{Effects of Arrival Processes}
Next, we are interested in understanding the effect of arrival processes on the capacity for the worst service time distribution. Specifically, we are interested in finding the arrival processes that maximize and minimize the capacity.

Analogous to Corollary \ref{cor:GeoG1Thre0} and Proposition \ref{prop:GeoG1Thre1} for
Type I systems, we have the following results.

\begin{corollary}
\label{cor:GxG1Thre0}
Consider the queue with given arrival rate $\lambda$ and service distribution $p_S$. If $H(\psi_0)<H(\psi_1)=H(\psi_2)=\cdots$, the capacity of the queue with Type II arrival is the same for all arrival distributions.
\end{corollary}

\begin{proposition}
\label{prop:GxGeo1Thre1Min}
For $H(\psi_0)=H(\psi_1)<H(\psi_2)=H(\psi_3)=\cdots$, for a given arrival rate $\lambda$ and a service distribution $p_S$, the capacity of the queue-channel over all Type II arrival processes with finite support $\{0, 1, \ldots, B\}$ is lower-bounded by $C_L = \lambda \left( \log|\mathbb{F}| + (H(\psi_2)-H(\psi_0)) (1-\frac{\lambda}{\mu})\frac{1}{k_0} - H(\psi_2) \right)$, where $k_0 = \sum_t (1-\frac{1}{B\lambda})^t p_S(t)$.
\end{proposition}
\begin{IEEEproof}
By an argument similar to the proof of Proposition \ref{prop:GeoG1Thre1}, the minimum is obtained when $k_0$ is maximized. Hence, it is sufficient to show the maximum value of $k_0$, thus the maximum of $m_A(0)$.

Towards this, we first show that the distribution
\begin{align*}
m_A^*(t) = \begin{cases} 1-\frac{1}{B\lambda}, & t=0 \\ \frac{1}{B\lambda}, & t=B \end{cases}
\end{align*}
maximizes $m_A(0)$ among all discrete distributions with bounded support $\{0,\ldots,B\}$ and mean $1/\lambda$.

This can be proved by contradiction. Suppose there is another distribution $m'_A$ with mean $1/\lambda$ and $m'_A(0) > m_A^*(0)$. Now 
\begin{align*}
\EX_{m'_A}[X] & = \sum_{t=0}^B t m'_{A}(t) \le B \sum_{t=1}^B m'_{A}(t)  = B (1-m'_A(0)) < B (1-m_A^*(0)) = \frac{1}{\lambda}, 
\end{align*}
which contradicts the assumption that $m'_A$ has expectation $1/\lambda$. Hence, there exists no $m'_A$ on $\{0,\ldots,B\}$ with $m'_A(0) > m_A^*(0)$.

Although the maximal $k_0$, $k_0^*$, is attained by $m_A^*(t)$, the induced Markov chain $Q$ is not irreducible because $m_A^*(1) = 0$. Instead, we use the approximate probability mass function $\tilde{m}_A(t)$, which has a nonzero mass at $t=1$. Define
\begin{align*}
\tilde{m}_A(t) = \begin{cases} 1-\frac{1}{B\lambda} - \epsilon \left( 1-\frac{1}{B\lambda} \right), & t=0 \\ \epsilon, & t=1 \\ \frac{1}{B\lambda} - \frac{\epsilon}{B\lambda}, & t=B. \end{cases}
\end{align*}
Then, we need to show $\tilde{m}_A(t)$ approximates $k_0$ arbitrarily close to $k_0^*$. Note that $k_0=\sum_t (m_A(0))^t p_S(t)$, which is a continuous function of $m_A(0)$. Thus, the conclusion follows.

Finally, the lower bound of capacity is computed as in the proof of Proposition \ref{prop:GeoG1Thre1},
\begin{align*}
C_L = \lambda \left( \log|\mathbb{F}| + (H(\psi_2)-H(\psi_0)) (1-\frac{\lambda}{\mu})\frac{1}{k_0} - H(\psi_2) \right),
\end{align*}
where $k_0 = \sum_t (1-\frac{1}{B\lambda})^t p_S(t)$.
\end{IEEEproof}

\begin{proposition}
\label{prop:GxGeo1Thre1Max}
For $H(\psi_0)=H(\psi_1)<H(\psi_2)=H(\psi_3)=\cdots$, for a given arrival rate $\lambda$ and a service distribution $p_S$,  the maximum capacity of the queue-channel over all Type II arrival processes with finite support $\{0, 1, \ldots, B\}$ is 
$C_U = \lambda \left( \log|\mathbb{F}| + (H(\psi_2)-H(\psi_0)) (1-\frac{\lambda}{\mu})\frac{1}{k_0} - H(\psi_2) \right)$, where $k_0 = \sum_t (1-\frac{1}{\lambda})^t p_S(t)$, attained by the Bernoulli arrival process, i.e.\, $m_A(0)=1-1/\lambda$ and $m_A(1)=1/\lambda$.
\end{proposition}
\begin{IEEEproof}
By a similar argument as above, the maximum is obtained when $k_0$ is minimized. This is reached when $m_A(0)$ is minimum. Hence, it is sufficient to prove that among all discrete distributions, Bernoulli achieves it.

Again, the proof is by contradiction. Let us assume there is another distribution $m'_A$ with the same mean, i.e.\, $\sum_t t m'_A(t) = 1/\lambda$ for which $m'_A(0)<m_A(0)$.
\begin{align*}
\sum_t t m'_A(t) \ge \sum_{t\ge 1} m'_A(t) = (1-m'_A(0)) > (1-m_A(0)) = \frac{1}{\lambda},
\end{align*}
which is a contradiction.

The capacity expression follows by substituting in the expression for $k_0$ for Bernoulli arrival.
\end{IEEEproof}
This proposition implies that having at most one arrival per time slot is better. In other words, burstiness in the arrival process hurts performance.

\section{Without Timing Information}
\label{sec:noTiming}
So far, we have assumed that the received or processed jobs have timestamps on dispatch time and completion time. Though this assumption is valid in many wireless settings (MAC timestamps are part of the protocols) and crowdsourcing scenarios (e.g., Samasource maintains timestamps), this information may not always be available. In this section we study the setting where the decoder does not have knowledge of $A^n,D^n$. 

Here, the decoder no longer observes $(Y^n,A^n,D^n)$, but only observes $Y^n$. Using the information spectrum technique it immediately follows that the capacity is
$$C(p_A) = \lambda \sup_{\PR(\mbf{X})} \underline{\mathbf{I}} (\mbf{X};\mbf{Y}).$$
The following theorem characterizes the capacity of the system based on the queue parameters and noise distributions.
Proof follows similar steps as the proof of Theorem \ref{thm:queueCapacity}.

\begin{theorem}
\label{thm:queueCapacityNoCSIR}
For a given arrival (dispatch) process distribution $p_A$ (or $m_A$) with $\lambda < \mu < 1$  which follows the assumption in Sec. \ref{sec:model}, there exists a distribution $\pi$ such that $\pi(q)>0$ for all $q \in \{0,1,\ldots\}$ and $\PR(Q_n) \to \pi$ as $n\to \infty$. The capacity of this queue-channel is $\lambda(\log |\mathbb{F}| - H(\sum_q \pi_q \psi_q))$, where $\pi_q \psi_q$ is a mixture of distributions $\{\psi_q\}$.
\end{theorem}
\begin{IEEEproof}
The converse is due to Fano's inequality. As the system satisfies the Markov relation $M-X^n-Y^n-\hat{M}$, using standard information-theoretic inequalities it follows that
\begin{align}
n\widetilde{R} \le \sum_{i=1}^n (H(X_i) - H(Y_i|X_i)) + n \epsilon. \nonumber 
\end{align} 
Note that $Y_i=X_i+Z_i$, and $Z_i$ is independent of everything else given $Q_i$. Thus, the distribution of $Z_i$ only depends on the distribution of $Q_i$. Also, note that $X_i$ is independent of the dispatch and service processes, and so is independent of $Q_i$. This implies the distribution of $Z_i$ is independent of $X_i$. 
So,
\begin{align}
H(Y_i|X_i) 
& = H(X_i + Z_i | X_i) = \sum_{x \in \mathbb{F}} \PR(X_i=x) H(Z_i + x|X_i=x) = H(Z_i). \nonumber
\end{align}

Note that $Z_i \sim \sum_q \PR(Q_i=q) \psi_q$. So, $H(Z_i) = H(\sum_q \PR(Q_i=q) \psi_q)$. The remainder of the proof follows since $\PR(Q_i) \to \pi$ and $\{Q_i\}$ is ergodic.

For achievability, like the proof of Theorem \ref{thm:queueCapacity} we pick a uniform and i.i.d.\ $P(X^n)$ and show that $\underline{\mathbf{I}}(\mbf{X};\mbf{Y})$ is equal to the expression in the theorem. Ergodicity of $\{Q_i\}$ implies that of $\{Z_i\}$ which is used to take the limit (almost surely) to evaluate $\underline{\mathbf{I}}(\mbf{X};\mbf{Y})$.
\end{IEEEproof}

Next, we consider some queuing systems to find the best dispatch and service processes in this setting. In the case of thresholded noise behavior the following result hold for a $G/\mathsf{geo}/1$ system.

\begin{proposition}
\label{prop:GGeo1NoCSIRMaxMin}
For a $G/\geo/1$ system with no timestamps, $\mathbb{F}=\{0,1\}$, $\PR(Z_i=1|q)\le 0.5$ for all $q$, and $H(\psi_0) = \cdots = H(\psi_b) < H(\psi_{b+1})=\cdots$ for some $b\in \{0,1,\ldots\}$, for a given $\lambda<\mu$ the queue-channel capacity is maximized by deterministic inter-arrival (for $\frac{1}{\lambda} \in \Z_+$) and is minimized by geometric inter-arrival among the class of sum-of-geometric random variables.
\end{proposition}
\begin{IEEEproof}
In this system $$H\left(\sum_q \pi_q \psi_q\right) = H\left(\psi_0 \sum_{q\le b} \pi_q + \psi_{b+1} \sum_{q\ge b+1} \pi_q\right).$$ Note that $\pi_q = (1-\sigma)\sigma^q$ where $\sigma$ is the fixed-point solution in Theorem \ref{thm:GGeo1}. Hence,
\begin{align*}
H\left(\sum_q \pi_q \psi_q\right) = H\left( \psi_0 (1-\sigma^{b+1}) + \psi_{b+1} \sigma^{b+1} \right).
\end{align*}
Now, as $\PR(Z_i=1|q)\le 0.5$, by monotonicity of binary entropy over $[0,0.5]$, it follows that the above expression is maximized (minimized) when $\sum_{q\ge b+1} \pi_q$ is maximized (minimized), which in turn happens when $\sigma$ is maximized (minimized).

The remainder of the argument follows as in the proof of Proposition \ref{prop:GGeo1Min} and \ref{prop:GGeo1Max}, because for $G/\geo/1$, deterministic arrival minimizes $\sigma$, while geometric arrival maximizes among the class of sum-of-geometric random variables.
\end{IEEEproof}

\begin{proposition}
\label{prop:GeoG1NoCSIRMaxMin}
For a $\geo/G/1$ system with no timestamps, $\mathbb{F}=\{0,1\}$, $\PR(Z_i=1|q)\le 0.5$ for all $q$, and $H(\psi_0) = H(\psi_1) < H(\psi_2) = H(\psi_{3}) = \cdots$, for a given $\lambda<\mu$ the queue-channel capacity is maximized by a deterministic service time and  (for $\frac{1}{\lambda} \in \Z_+$) and is minimized by geometric service time among the class of sum-of-geometric random variables. 
\end{proposition}
\begin{IEEEproof}
By the same argument as in proof of Proposition \ref{prop:GGeo1NoCSIRMaxMin}, the maximum is achieved when $\pi_0+\pi_1$ is maximized. The remaining argument follows the proof of Proposition \ref{prop:GeoG1Thre1}.
\end{IEEEproof}

\section{Conclusion}
\label{sec:timingCapacity}
Inspired by several engineering applications, we studied the performance of servers with queue-length dependent service quality using an information-theoretic approach. We defined the capacity of such queuing systems to be the maximum rate at which jobs can be processed
with arbitrarily small error probability, and characterized it in terms of queuing parameters.

We studied Type I and Type II arrivals separately for technical reasons. In Type I arrivals with some assumptions, for the $G/\geo/1$ queue, jobs arriving deterministically maximize capacity while bursty arrivals minimize capacity.  Similarly, for the $\geo/G/1$ queue, deterministic service maximizes capacity, but bursty service minimizes capacity. Type II arrivals give similar conclusions except that Bernoulli arrivals maximize capacity for the $G/\geo/1$ queue.

\begin{appendix}

\subsection{Proof of Lemma~\ref{lem:Qstationary}}
\label{app:lem_qstationary}
We need separate approaches for Type I and Type II arrival processes, as the nature of $\{Q_i\}$ process depends on the type. 

First consider Type II processes. For these, $\lambda < 1$ implies $\sum_k k m_A(k) < 1$. If $m_A(0) = 0$, the mean arrival rate must be equal or greater than $1$, contradicting the assumption. Hence, $m_A(0)>0$. Also from the assumption in Sec.~\ref{sec:model}, $m_A(1)>0$.

Under the assumptions, we show $\{Q_i\}$ is an irreducible and aperiodic Markov chain by proving $\PR(Q_{i+1}=Q_i+1|Q_i), \PR(Q_{i+1}=\max(Q_i-1,0)|Q_i), \PR(Q_{i+1} = Q_i|Q_i) > 0$ for all $Q_i$. If this is true then any state can be reached from any other state, since states are in $\Z_+$.
Notice the enumerated probabilities are probabilities of the events corresponding to two, one, and no arrivals, respectively, during a service time. 

By the above result and assumption, $m_A(0), m_A(1)>0$ and there exists an $s>1$ such that $p_S(s)>0$. Note the probability of exactly two arrivals in a service time is lower bounded by $$(m_A(1))^2 (m_A(0))^{s-2} p_S(s)$$ for any $s>1$. As there exists an $s>1$ such that $p_S(s)>0$ and $m_A(a), m_A(1)>0$, this bound is strictly positive. Probability of exactly one arrival in a service time is lower bounded by $m_A(1) (m_A(0))^{s-1} p_S(s)$ for any $s>0$, which again is strictly positive. Probability of no arrival is lower bounded by $p_S(s) (m_A(0))^{s} $, which is also strictly positive.

Note that as $\PR(Q_{i+1}=Q_i| Q_i)>0$, this Markov chain is also aperiodic. Due to the self-loop if $\PR(Q_{i+k}=q'|Q_i=q)$ is positive, then so is $\PR(Q_{i+k+1}=q'|Q_i=q)$.

Positive recurrence follows by considering queue-length to be the Lyapunov function, because $\lambda < \mu$. Hence, the result follows for Type II processes due to the existence of a unique stationary distribution for an irreducible and aperiodic positive recurrent Markov chain. Hence, $\{Q_i\}$ is ergodic.

For Type I, $\{Q_i\}$ is not a Markov chain, and we take a different approach.
First note that as $\mu<1$, $\sum_s s p_S(s) > 1$. This implies there exists an $s>1$ such that $p_S(s)>0$. Note that by assumption $\lambda<\mu$. Then, for Type I arrival processes this implies there exists an $a>1$ such that $p_A(a)>0$.

Consider the process $\{W_i\}$, the sojourn time for jobs. We first claim that under the assumption, this is an irreducible, aperiodic, and positive recurrent Markov chain. It is known in queuing theory that for i.i.d.\ inter-arrival and service times, $\{W_i\}$ is a Markov chain. Next, we show irreducibility and aperiodicity by showing that $\PR(W_{i+1}=W_i+1|W_i)$, $\PR(W_{i+1}=W_i|W_i)$, and $\PR(W_{i+1}=\max(W_i-1,0),W_i) > 0$.

First, we consider the case when $p_A$ has a support that spans $\Z_+$.
As $\mu<1$, there exists an $s>1$ such that $p_S(s)>0$. 
Consider a possible path from $W_i$ to $W_{i+1}=\max(W_i+b,0), b \in \{0,\pm 1\}$. This can happen as follows: the $(i+1)$th job brings a service time requirement of $s$, and it reaches the system $s-b$ time after the $i$th job. As the service times and inter-arrival times are independent, probability of this sample path event is exactly $p_S(s) p_A(s-b)$, which is strictly positive.

Next, we consider the case when $p_S$ has a support spanning $\Z_+$. As
$\lambda<1$, there exists an $a>1$ such that $p_A(a)>0$. Then a possible path for the events is as follows: the $(i+1)$th job comes $a$ time after $i$th job and brings with it a service requirement of $a+b$. The rest follows by evaluating the probability of this event.

Note that $\{W_i\}$ is an irreducible and aperiodic Markov chain. Note that given $W_i$, $Q_i$ is independent of anything else because given $W_i$, it only depends on the number of arrivals in the time $W_i$: 
$$\PR(Q_i=q) = \PR \left(\sum_{i=1}^q A_i \le W_i < \sum_{i=1}^{q+1} A_i\right).$$
As the $A_i$ are i.i.d., this also implies that given a distribution of $W_i$, the distribution of $Q_i$ is fixed. 

It follows from queuing theory that $\{W_i\}$ is positive recurrent for $\lambda<\mu$. Hence, $\{W_i\}$ converges in distribution to a stationary distribution, and by the above argument, so does $\{Q_i\}$. Ergodicity of $Q_i$ follows from the ergodicity of $W_i$. 

$\blacksquare$

\subsection{Proof of Theorem \ref{thm:GGeo1}}
\label{app:pf_thm2}
Let $\{\hat{Q}_i\}$ be queue-lengths seen by the arrivals, then the stationary distribution of $\hat{Q}_i$ is the same as that of $Q_i$. Note that there is only one arrival and one departure at a time. Since the queue-length is stable, the fraction of time the queue-length increases by $1$ from a value $q$ is the same as the fraction of time the queue-length decreases by $1$ from $q$, for all $q$. Since increase corresponds to arrival and decrease corresponds to departure, the fraction of arrivals and departures that see a queue-length $q$ is the same. Thus it is sufficient to show that the stationary distribution of $\{\hat{Q}\}$ is $\pi_k = (1-\sigma)\sigma^k$, where $\sigma$ solves $x = \sum_{n=0}^\infty p_A(n) (1-\mu+x\mu)^n$ in $(0,1)$.

We shall first show the uniqueness of the stationary distribution from the fact that $\{\hat{Q}_i\}$ is an irreducible Markov chain, and then derive the stationary distribution.

Consider the transition probability
\begin{align}
& \PR(\hat{Q}_{i+1} = q' | \hat{Q}_i = q, \hat{Q}_{i-1}, \ldots). \nonumber
\end{align}
As at most one arrival is possible, the probability is $0$ for 
$q'-q>1$. For $q'-q\le 1$,
\begin{align}
& \PR(\hat{Q}_{i+1} = q' | \hat{Q}_i = q, \hat{Q}_{i-1}, \ldots) \nonumber \\
&= \PR(\mbox{there are} \ q-q'+1 \ \mbox{departures between $i$ and $i+1$ arrival} ~|~ \hat{Q}_i = q, \hat{Q}_{i-1}, \ldots)\mbox{.} \nonumber
\end{align}
As service time is geometric with mean $\frac{1}{\mu}$ and hence memoryless, starting from any time, the time to the next departure is geometric with the same mean, if there is a job in the queue. After any arrival, there is always at least one job in the queue, and hence, time to the next departure is geometric. Thus the probability that there are
$q-q'+1$ departures given the past is nothing but the probability that the sum of $q-q'+1$ geometric random variables is less than a realization of $p_A$. Thus,
\begin{align}
\PR(\hat{Q}_{i+1} = q' | \hat{Q}_i = q, \hat{Q}_{i-1}, \ldots) &= \sum_{t=0}^\infty p_A(t) \PR\left(\sum_{i=1}^{q-q'+1} S_i \le t \le \sum_{i=1}^{q-q'} S_i \right) \nonumber \\
& = \sum_{t=0}^\infty p_A(t) \PR(\mbox{Bin}(t, \mu) = q-q'+1) \label{eq:GGeo1Tran1}  \\
& = \sum_{t=0}^\infty p_A(t) {t \choose q-q'+1} (1-\mu)^{t-q+q'-1} \mu^{q-q'+1}. \nonumber
\end{align}

Eq.~\eqref{eq:GGeo1Tran1} follows because the service times are geometric, meaning each time a job in service gets completed according to a Bernoulli random variable, and the sum of Bernoulli random variables is binomial. This derivation implies the transition depends only on $q$ and $q'$, further implying the process is Markov.

Thus the probability of the $q \to 0$ transition is
$$\sum_{t=0}^\infty p_A(t) {t \choose q+1} (1-\mu)^{t-q-1} \mu^{q+1}.$$
Note that the transitions can be written as the amount of change in the queue-length, meaning a $q \to q'$ transition is a $q-q'$ change, and is nothing but the probability of having $q-q'+1$ departures before an arrival.

For $k \ge 0$, let $\beta_k$ denote the probability that the sum of $k$  geometric random variables is less than the time between two arrivals. Then, for $q'>0$,
$$\PR(\hat{Q}_{i+1} = q' | \hat{Q}_i = q) = \beta_{q-q'+1},$$
and for $q'=0$,
$$\PR(\hat{Q}_{i+1} = 0 | \hat{Q}_i = q) = 1 -  \sum_{k=0}^q \beta_k.$$

Also, as $\beta_0, \beta_1, \beta_2 >0$, the Markov chain is irreducible and aperiodic. Thus there exists a unique stationary distribution $\pi$ which solves $\pi = \pi [P]$, where $[P]$ is the probability transition matrix. The transition matrix $[P]$ is written as a matrix whose first column is $(1 - \beta_0, 1 -  \sum_{k=0}^1 \beta_k, \ldots)^T$ and other columns are $(0, \ldots, \beta_0, \beta_1, \beta_2, \ldots)^T$, where $\beta_0$ is the $(i,i+1)$th entry.

From $\pi = \pi [P]$ it follows that 
\begin{align}
\pi_0 & = \sum_{i=0}^\infty \left(1-\sum_{k=0}^i \beta_k \right) \pi_i, \nonumber \\
\pi_k & = \sum_{i=0}^\infty \pi_{k-1+i} \beta_i ~~~ \mbox{for } k>0. \nonumber
\end{align}
Like in the analysis of $GI/M/1$ queue \cite{Kleinrock1975_I}, we guess a solution
$\pi_k = \pi_0 \sigma^k$ for some $\sigma<1$. Next, we check if this solution satisfies $\pi = \pi [P]$ for a unique $\sigma<1$.

It follows from $\pi=\pi [P]$, as above, that $\sigma$ must satisfy
\begin{align}
\sigma & = \sum_{i=0}^\infty \sigma^i \beta_i = \sum_{i=0}^\infty \sigma^i \sum_{t=0}^\infty p_A(t) {t \choose i} (1-\mu)^{t-i} \mu^{i} \nonumber \\
& = \sum_{t=0}^\infty p_A(t) \sum_{i=0}^t {t \choose i} (1-\mu)^{t-i} (\sigma \mu)^{i} \label{eq:sigma1} \\
& = \sum_{t=0}^\infty p_A(t) (1-\mu+\sigma \mu)^t. \label{eq:sigma2}
\end{align}
Eq.~\eqref{eq:sigma1} follows by interchanging the two sums, as per the Fubini-Tonelli theorem since terms are non-negative. Eq.~\eqref{eq:sigma2} follows using the binomial theorem.

To show that the distribution $\pi$ is unique, we show that
$x=\sum_{t=0}^\infty p_A(t) (1-\mu+x \mu)^t$ has a unique solution in $0<x<1$. 
Towards this we characterize $\sum_{t=0}^\infty p_A(t) (1-\mu+x \mu)^t$, in Lemmas~\ref{lem:xConvex} and \ref{lem:uniqueSigma} given below, which complete the proof. $\blacksquare$

\begin{lemma}
\label{lem:xConvex}
For any $p_A$ on $\Z_+$ and $\mu \in (0,1)$, $\sum_{t=0}^\infty p_A(t) (1-\mu+x \mu)^t$ is an increasing function of $x$ in $(0,1)$, and strictly convex in $(0,1)$.
\end{lemma}
\begin{IEEEproof}
Define 	
\[
f(x) = \sum_{t=0}^\infty p_A(t) (1-\mu+x \mu)^t.
\]
It is sufficient to show that $f'(x), f''(x)$ are both strictly positive in $x \in (0,1)$.

Let the partial sum up to $T$ in $f(x)$ be $f_T(x)$, i.e.\,
\begin{align*}
f_T(x) &= \sum_{t=0}^T p_A(t) (1-\mu+x \mu)^t,
\end{align*}
and then 
\begin{align*}
f_T'(x) &= \sum_{t=0}^T \mu t p_A(t) (1-\mu+x \mu)^t.
\end{align*}
It is easy to see that $f_T'(x)$ is increasing as $0 < 1-\mu+x \mu < 1$. In addition, $f_T'(x)$ is bounded since
\begin{align*}
f_T'(x) &= \sum_{t=0}^T \mu t p_A(t) (1-\mu+x \mu)^t \le \mu \sum_{t=0}^T t p_A(t) < \infty.
\end{align*}
Since $f_T'(x)$ is increasing and bounded, $\lim_{T\to \infty} f_T'(x)$ exists for all $x \in (0,1)$.

Next, note that for any $x \in (0,1)$, the difference between $f_T'(x)$ and $f'(x)$ is
\begin{align*}
f'(x) - f_T'(x) &= \sum_{t=T+1}^{\infty} \mu t p_A(t) (1-\mu+x \mu)^{t-1} \le \mu \sum_{t=T+1}^{\infty} t p_A(t) \to 0
\end{align*}
as $T \to \infty$, where the inequality follows from $0 < 1-\mu+x \mu < 1$ and the limit follows from the condition of fixed mean. Then, $\lim_{T\to \infty} f_T'(x) = f'(x)$ uniformly in $(0,1)$. That $f'(x) > 0$ follows from 
\begin{equation*}
0 < 1-\mu+x \mu < 1.
\end{equation*}

Similarly, we can show the existence and strict positivity of $f''(x)$, which completes the proof.
\end{IEEEproof}

\begin{lemma}
\label{lem:uniqueSigma}
The equation $x=\sum_{t=0}^\infty p_A(t) (1-\mu+x \mu)^t$ has a unique solution in $(0,1)$.
\end{lemma}
\begin{IEEEproof}
Note that $x=1$ is a solution to this fixed-point equation. First, we show that there is at least one fixed point in $(0,1)$. 

Again, $f(x)=\sum_{t=0}^\infty p_A(t) (1-\mu+x \mu)^t > 0$ for $x=0$. Hence, if there is no fixed point in $(0,1)$ this implies that $f(x)$ is
strictly greater than $x$ in $(0,1)$. 

Now, consider the  derivative of $f(x)$ at $1-\frac{\delta}{\mu}$, 
This is $\mu \sum_t t (1-\delta)^t p_A(t) = \mu \hat{A}(1-\delta)$, where $\hat{A}(\alpha) = \EX_{p_A}\alpha^A$. We know that generating function $\hat{A}$ is continuous around $1$. Hence, as $\delta \to 0$, $\hat{A}(1-\delta) \to \frac{1}{\lambda}$. As $\frac{\mu}{\lambda}>1$, there exists $\delta>0$ such that $\mu \sum_t t (1-\delta)^t p_A(t) > 1$. This means that the derivative of $f(x)$ at $x=1-\delta$ is $>1$. 

If $f(x) > x$ for all $x \in (0,1)$, then the following is true. 
From convexity of $f$,
\begin{align}
f(1) &\ge f(1-\delta)  + \delta f'(1-\delta) > 1-\delta + \delta f'(1-\delta) > 1-\delta+\delta = 1. \nonumber
\end{align}
This is a contradiction. So, there exists a fixed point in $(0,1)$.

Let us assume there are more than one fixed points in $(0,1)$. By Lemma \ref{lem:xConvex}, $f(x)$ is convex in $(0,1)$. A convex function can intersect a line at most twice. As $f(x)$ crosses $y=x$ at $x=1$, there can be only one fixed point in $[0,1)$, but $0$ is not a fixed point.
\end{IEEEproof}

\subsection{Proof of Lemma \ref{lem:geoProdSum}}
\label{app:pf_lem_prodsum}
Note that for a geometric random variable $A_i$ with mean $1/\lambda_i$ and letting $\alpha = (1-\mu-\sigma\mu) \in (0,1)$,
\begin{align}
\tilde{A}(p_{A_i},\sigma) &= \sum_{t=1}^\infty \alpha^t p_{A_i}(t) = \sum_{t=1}^\infty \alpha^t (1-\lambda_{i})^{t-1} \lambda_{i} = \alpha \lambda_{i} \sum_{t=0}^\infty (\alpha(1-\lambda_{i}))^t \nonumber \\
&= \frac{\alpha \lambda_{i}}{(1-\alpha)+\alpha\lambda_{i}} = \frac{\frac{\alpha}{1-\alpha}} {\frac{1}{\lambda_{i}} + \frac{\alpha}{1-\alpha}} \mbox{.} \label{eq:geo_charac}
\end{align}

Consider the sum-of-geometric random variables $\mathcal{A}^s$ first. Then for any sum-of-geometric random variable $p_{A^s} \in \mathcal{A}^s$, 
\begin{align*}
\tilde{A}(p_{A^s}, \sigma) &= \sum_{t=1}^\infty \alpha^t p_{A^s}(t) = \sum_{t_i=1, \ 1\le i \le I}^\infty \alpha^{t_1+t_2+\cdots+t_I} p_{A_1}(t_1) \cdots p_{A_I}(t_I) \\
&= \prod_{i=1}^I \sum_{t_i=1}^{\infty} \alpha^{t_i} p_{A_i}(t_i) = \prod_{i=1}^I \frac{\frac{\alpha}{1-\alpha}} {\frac{1}{\lambda_{i}} + \frac{\alpha}{1-\alpha}} \mbox{.}
\end{align*}
The last equality follows from \eqref{eq:geo_charac}.
Note that the inequality $$\prod_i (1+x_i) \ge 1+\sum_i x_i$$ holds for any $x_i > 0$.
Hence, inverting both sides of this inequality and scaling both numerator and denominator by $\frac{\alpha}{1-\alpha}$,
\begin{align*}
\tilde{A}(p_{A^s}, \sigma) &= \prod_{i=1}^I \frac{\frac{\alpha}{1-\alpha}} {\frac{1}{\lambda_{i}} + \frac{\alpha}{1-\alpha}} \le \frac{\frac{\alpha}{1-\alpha}} {\sum_{i=1}^I \frac{1}{\lambda_i} + \frac{\alpha}{1-\alpha}} = \frac{\frac{\alpha}{1-\alpha}} {\frac{1}{\lambda} + \frac{\alpha}{1-\alpha}} \\
&= \tilde{A}(\geo, \sigma).
\end{align*}

Next since $A^m$ is mixed, for any $p_{A^m} \in \mathcal{A}^m$, 
\begin{align*}
\tilde{A}(p_{A^m}, \sigma) &= \sum_{t=1}^\infty \alpha^t p_{A^m}(t) = \sum_{i=1}^I c_i \sum_{t_i=1}^\infty \alpha^{t_i} p_{A_i}(t_i) = \sum_{i=1}^I c_i \frac{\frac{\alpha}{1-\alpha}} {\frac{1}{\lambda_{i}} + \frac{\alpha}{1-\alpha}}.
\end{align*}
The last expression is convex in $1/\lambda_i$. Hence by Jensen's inequality
\begin{align*}
\tilde{A}(p_{A^m}, \sigma) &= \sum_{i=1}^I c_i \frac{\frac{\alpha}{1-\alpha}} {\frac{1}{\lambda_{i}} + \frac{\alpha}{1-\alpha}} \ge \frac{\frac{\alpha}{1-\alpha}} {\sum_{i=1}^I c_i \frac{1}{\lambda_{i}} + \frac{\alpha}{1-\alpha}} = \frac{\frac{\alpha}{1-\alpha}} {\frac{1}{\lambda} + \frac{\alpha}{1-\alpha}} \\
&= \tilde{A}(\geo,\sigma).
\end{align*}

\subsection{Proof of Theorem \ref{thm:GeoG1}}
\label{app:pf_thm3}
Consider the following transition probability for $q>0$.
\begin{align}
&\PR(Q_{i+1}=q'|Q_i=q, Q_{i-1}, \ldots) \nonumber \\
& = \PR(\mbox{there are $q'-q+1$ arrivals between departures $i-1$ and $i$} ~|~ Q_i=q, Q_{i-1}, \ldots)\nonumber \\
& = \PR(\mbox{sum of $q'-q+1$ geometric times} \le \mbox{interdeparture time between $i-1$ and $i$}) \label{eq:GeoG1Tran1} \\
& = \sum_{t=0}^\infty p_S(t) \PR(\mbox{Bin}(t,\lambda) = q'-q+1) \label{eq:GeoG1Tran2} \\
& = \sum_{t=0}^\infty p_S(t) {t \choose q'-q+1} (1-\mu)^{t-q'+q-1} \mu^{q'-q+1} \nonumber \\
& = k_{q'-q+1} \nonumber
\end{align}
Eq.~\eqref{eq:GeoG1Tran1} follows because geometric random variables are memoryless. Geometric inter-arrival is the same as Bernoulli arrival per time slot, and the sum of Bernoulli variables is binomial, which leads to \eqref{eq:GeoG1Tran2}. 

When $Q_i=0$, note that just before the $(i+1)$th arrival, the queue-length is $0$, and it is $1$ just after the $(i+1)$th arrival. Then the probability that $Q_{i+1}=q'$ is equal to the probability that there are exactly $q'$ arrivals during the service time of the $(i+1)$th job. From above, this is equal to $k_{q'}$.

This proves $\{Q_i\}$ is Markov; irreducibility and aperiodicity follows since $\PR(Q_{i+1}=Q_i+\delta|Q_i) > 0$ for $\delta \in \{0, \pm 1\}$.

From $\pi = \pi [P]$ for this Markov chain it follows that
\begin{align}
\pi_0 k_0 + \pi_1 k_0 &= \pi_0 \nonumber \\
\pi_0 k_1 + \pi_1 k_1 + \pi_2 k_0 &= \pi_1 \nonumber \\
&\vdots \nonumber
\end{align}
Multiplying the first equation by $z^0$, the second by $z$, the third by $z^2$, and so on, and then summing all of them we get
\begin{align}
& \pi_0 K(z) + K(z)(\pi_1 + \pi_2 z + \cdots) = \Pi(z), \nonumber
\end{align}
which, after some algebra, gives
\begin{align}
\Pi(z) = \frac{\pi_0(z-1)K(z)}{z-K(z)}. \nonumber
\end{align}
We know that $\Pi(1)=1$, then the left side must also be $1$ for $z=1$. But it is $\frac{0}{0}$ when evaluated at $z=1$, as $K(1)=\sum_j k_j = 1$. Thus using l'H\^{o}pital's rule we get
$$\pi_0 = \frac{1-K'(1)}{K(1)}.$$

Note that $K'(z) = \sum_j j k_j z^j$ which gives $K'(1)=\sum_j j k_j$, i.e.\, $K'(1)$ is the expected number of arrivals in a time distributed as $p_S$. As arrivals are Bernoulli and are independent from service times, from Wald's lemma we get 
$$K'(1) = \frac{\lambda}{\mu},$$
which in turn gives $\pi_0 = 1-\frac{\lambda}{\mu}$.

From $\Pi(z)$ we can obtain $\pi_1$ by evaluating 
$\frac{\Pi(z)-\pi_0}{z}$ as $z\to 0$. By repeating the procedure we can obtain $\pi_k$ by evaluating the limit of 
$\frac{\Pi(z) - \sum_{j=0}^{k-1}\pi_j z^j}{z^k}$ as $z \to 0$.

\end{appendix}
\bibliographystyle{IEEEtran}
\bibliography{conf_abrv,abrv,lrv_lib}

\newcommand{\SortNoop}[1]{}
\begin{thebibliography}{10}
\providecommand{\url}[1]{#1}
\csname url@samestyle\endcsname
\providecommand{\newblock}{\relax}
\providecommand{\bibinfo}[2]{#2}
\providecommand{\BIBentrySTDinterwordspacing}{\spaceskip=0pt\relax}
\providecommand{\BIBentryALTinterwordstretchfactor}{4}
\providecommand{\BIBentryALTinterwordspacing}{\spaceskip=\fontdimen2\font plus
\BIBentryALTinterwordstretchfactor\fontdimen3\font minus
  \fontdimen4\font\relax}
\providecommand{\BIBforeignlanguage}[2]{{%
\expandafter\ifx\csname l@#1\endcsname\relax
\typeout{** WARNING: IEEEtran.bst: No hyphenation pattern has been}%
\typeout{** loaded for the language `#1'. Using the pattern for}%
\typeout{** the default language instead.}%
\else
\language=\csname l@#1\endcsname
\fi
#2}}
\providecommand{\BIBdecl}{\relax}
\BIBdecl

\bibitem{EphremidesH1998}
A.~Ephremides and B.~Hajek, ``Information theory and communication networks: An
  unconsummated union,'' \emph{{IEEE} Trans. Inf. Theory}, vol.~44, no.~6, pp.
  2416--2434, Oct. 1998.

\bibitem{Schwartz1978}
B.~Schwartz, ``Queues, priorities, and social process,'' \emph{Social
  Psychology}, vol.~41, no.~1, pp. 3--12, Mar. 1978.

\bibitem{BransonVWPB2014}
S.~Branson, G.~Van~Horn, C.~Wah, P.~Perona, and S.~Belongie, ``The ignorant led
  by the blind: A hybrid human-machine vision system for fine-grained
  categorization,'' \emph{Int. J. Comput. Vis.}, vol. 108, no. 1-2, pp. 3--29,
  May 2014.

\bibitem{BorokhovichCRVV2015}
M.~Borokhovich, A.~Chatterjee, J.~Rogers, L.~R. Varshney, and S.~Vishwanath,
  ``Improving impact sourcing via efficient global service delivery,'' in
  \emph{Proc. Data for Good Exchange (D4GX)}, Sep. 2015.

\bibitem{VempatyVV2014}
A.~Vempaty, L.~R. Varshney, and P.~K. Varshney, ``Reliable crowdsourcing for
  multi-class labeling using coding theory,'' \emph{{IEEE} J. Sel. Topics
  Signal Process.}, vol.~8, no.~4, pp. 667--679, Aug. 2014.

\bibitem{SriramL1989}
K.~Sriram and D.~M. Lucantoni, ``Traffic smoothing effects of bit dropping in a
  packet voice multiplexer,'' \emph{{IEEE} Trans. Commun.}, vol.~37, no.~7, pp.
  703--712, Jul. 1989.

\bibitem{DraperTW2005}
S.~C. Draper, M.~D. Trott, and G.~W. Wornell, ``A universal approach to queuing
  with distortion control,'' \emph{{IEEE} Trans. Autom. Control}, vol.~50,
  no.~4, pp. 532--537, Apr. 2005.

\bibitem{Goyal2001b}
V.~K. Goyal, ``Multiple description coding: Compression meets the network,''
  \emph{{IEEE} Signal Process. Mag.}, vol.~18, no.~5, pp. 74--93, Sep. 2001.

\bibitem{AnantharamV1996}
V.~Anantharam and S.~{Verd\'{u}}, ``Bits through queues,'' \emph{{IEEE} Trans.
  Inf. Theory}, vol.~42, no.~1, pp. 4--18, Jan. 1996.

\bibitem{BedekarA1998}
A.~S. Bedekar and M.~Azizo{\~{g}}lu, ``The information-theoretic capacity of
  discrete-time queues,'' \emph{{IEEE} Trans. Inf. Theory}, vol.~44, no.~2, pp.
  446--461, Mar. 1998.

\bibitem{PrabhakarG2003}
B.~Prabhakar and R.~Gallager, ``Entropy and the timing capacity of discrete
  queues,'' \emph{{IEEE} Trans. Inf. Theory}, vol.~49, no.~2, pp. 357--370,
  Feb. 2003.

\bibitem{SundaresanV2000}
R.~Sundaresan and S.~Verd{\'{u}}, ``Sequential decoding for the exponential
  server timing channel,'' \emph{{IEEE} Trans. Inf. Theory}, vol.~46, no.~2,
  pp. 705--709, Mar. 2000.

\bibitem{WagnerA2005}
A.~B. Wagner and V.~Anantharam, ``Zero-rate reliability of the
  exponential-server timing channel,'' \emph{{IEEE} Trans. Inf. Theory},
  vol.~51, no.~2, pp. 447--465, Feb. 2005.

\bibitem{GilesH2002}
J.~Giles and B.~Hajek, ``An information-theoretic and game-theoretic study of
  timing channels,'' \emph{{IEEE} Trans. Inf. Theory}, vol.~48, no.~9, pp.
  2455--2477, Sep. 2002.

\bibitem{GongKV2011}
X.~Gong, N.~Kiyavash, and P.~Venkitasubramaniam, ``Information theoretic
  analysis of side channel information leakage in {FCFS} schedulers,'' in
  \emph{Proc. 2011 IEEE Int. Symp. Inf. Theory}, Jul. 2011, pp. 1255--1259.

\bibitem{GorantlaKKCMK2012}
S.~K. Gorantla, S.~Kadloor, N.~Kiyavash, T.~P. Coleman, I.~S. Moskowitz, and
  M.~H. Kang, ``Characterizing the efficacy of the {NRL} network pump in
  mitigating covert timing channels,'' \emph{{IEEE} Trans. Inf. Forensics
  Security}, vol.~7, no.~1, pp. 64--75, Feb. 2012.

\bibitem{TavanYB2013}
M.~Tavan, R.~D. Yates, and W.~U. Bajwa, ``Bits through bufferless queues,'' in
  \emph{Proc. 51st Annu. Allerton Conf. Commun. Control Comput.}, Oct. 2013,
  pp. 755--762.

\bibitem{Telatar1992}
{\.{I}}.~E. Telatar, ``Multi-access communications with decision feedback
  decoding,'' Ph.D.~thesis, Massachusetts Institute of Technology, Cambridge,
  MA, May 1992.

\bibitem{TelatarG1995}
{\.{I}}.~E. Telatar and R.~G. Gallager, ``Combining queueing theory with
  information theory for multiaccess,'' \emph{{IEEE} J. Sel. Areas Commun.},
  vol.~13, no.~6, pp. 963--969, Aug. 1995.

\bibitem{RajTT2004}
S.~Raj, E.~Telatar, and D.~Tse, ``Job scheduling and multiple access,'' in
  \emph{Advances in Network Information Theory}, P.~Gupta, G.~Kramer, and A.~J.
  van Wijngaarden, Eds.\hskip 1em plus 0.5em minus 0.4em\relax Providence:
  DIMACS, American Mathematical Society, 2004, pp. 127--137.

\bibitem{MusyT2006}
S.~Musy and E.~Telatar, ``On the transmission of bursty sources,'' in
  \emph{Proc. 2006 IEEE Int. Symp. Inf. Theory}, Jul. 2006, pp. 2899--2903.

\bibitem{MichelusiBEMM2015_arXiv}
N.~Michelusi, J.~Boedicker, M.~Y. El-Naggar, and U.~Mitra, ``Queuing models for
  abstracting interactions in bacterial communities,'' arXiv:1508.00942
  [cs.ET]., Aug. 2015.

\bibitem{Harris1967}
C.~M. Harris, ``Queues with state-dependent stochastic service rates,''
  \emph{Oper. Res.}, vol.~15, no.~1, pp. 117--130, Jan.-Feb. 1967.

\bibitem{Pinsker1964}
M.~S. Pinsker, \emph{Information and Information Stability of Random Variables
  and Processes}.\hskip 1em plus 0.5em minus 0.4em\relax San Francisco:
  Holden-Day, 1964.

\bibitem{VerduH1994}
S.~{Verd\'{u}} and T.~S. Han, ``A general formula for channel capacity,''
  \emph{{IEEE} Trans. Inf. Theory}, vol.~40, no.~4, pp. 1147--1157, Jul. 1994.

\bibitem{Han2003}
T.~S. Han, \emph{Information-Spectrum Methods in Information Theory}.\hskip 1em
  plus 0.5em minus 0.4em\relax Berlin: Springer, 2003.

\bibitem{CoverT1991}
T.~M. Cover and J.~A. Thomas, \emph{Elements of Information Theory}.\hskip 1em
  plus 0.5em minus 0.4em\relax New York: John Wiley \& Sons, 1991.

\bibitem{CaireS1999}
G.~Caire and S.~Shamai~(Shitz), ``On the capacity of some channels with channel
  state information,'' \emph{{IEEE} Trans. Inf. Theory}, vol.~45, no.~6, pp.
  2007--2019, Sep. 1999.

\bibitem{Kleinrock1975_I}
L.~Kleinrock, \emph{Queuing Systems, Volume I: Theory}.\hskip 1em plus 0.5em
  minus 0.4em\relax John Wiley \& Sons, Inc., 1975.

\bibitem{ChatterjeeVV2015}
A.~Chatterjee, L.~R. Varshney, and S.~Vishwananth, ``Work capacity of freelance
  markets: Fundamental limits and decentralized schemes,'' in \emph{Proc. 2015
  IEEE INFOCOM}, Apr. 2015, pp. 1769--1777.

\end{thebibliography}

\end{document}